\documentclass[preprint,secnumarabic,amssymb, nobibnotes, aps, pr,superscriptaddress]{revtex4-2}
\usepackage{subfigure}
\usepackage{amsmath}
\usepackage{graphicx}% Include figure files
\usepackage{dcolumn}% Align table columns on decimal point
\usepackage{bm}% bold math
\usepackage[utf8]{inputenc}
\usepackage{textcomp}
\usepackage{amsmath}
\usepackage[version=4]{mhchem}
\usepackage{siunitx}
\usepackage{caption}
\usepackage{soul}
\soulregister\cite7 % 针对\cite命令
\soulregister\citep7 % 针对\citep命令
\soulregister\citet7 % 针对\citet命令
\soulregister\ref7 % 针对\ref命令
\title{Adaptive extended Kalman filter and point ahead angle prediction in the detection of gravitational waves in space.}

\def\ot{\scriptscriptstyle{12}}

\begin{document}
	
	%\preprint{APS/123-QED}
	
	% Force line breaks with \\

	\title{Adaptive extended Kalman filter and point ahead angle prediction in the detection of gravitational waves in space.}

	\author{Jinke Yang}	\affiliation{Shanghai Institute of Technical Physics, Chinese Academy of Sciences, Shanghai, 200083, China.}
	\affiliation{%
		University of Chinese Academy of Sciences, Beijing, 100049, China.}%
	\author{Yong Xie}%
	\affiliation{%
		Shanghai Institute of Technical Physics, Chinese Academy of Sciences, Shanghai, 200083, China.}%
	\author{Wenlin Tang}%
	\affiliation{%
		Key Laboratory of Electronics and Information Technology for Space System, National Space Science Center,Chinese Academy of Sciences, Beijing 100190, China}%
	\author{Xindong Liang}%
	\affiliation{%
		Hangzhou Institute for Advanced Study, UCAS, Hangzhou 310024, China.}%
	\author{Liang Zhang}%
	\affiliation{%
		Shanghai Institute of Technical Physics, Chinese Academy of Sciences, Shanghai, 200083, China.}%
	\affiliation{%
		University of Chinese Academy of Sciences, Beijing, 100049, China.}%
	\author{Zhao Cui}%
	\affiliation{%
		University of Chinese Academy of Sciences, Beijing, 100049, China.}%
	\affiliation{%
		Hangzhou Institute for Advanced Study, UCAS, Hangzhou 310024, China.}%
	\author{Xue Wang}%
	\affiliation{%
		Shanghai Institute of Technical Physics, Chinese Academy of Sciences, Shanghai, 200083, China.}%
	\affiliation{%
		University of Chinese Academy of Sciences, Beijing, 100049, China.}%
	\author{Haojie Li}%
	\affiliation{%
		Hangzhou Institute for Advanced Study, UCAS, Hangzhou 310024, China.}%
	\author{Jianjun Jia}%
	\email{jjjun10@mail.sitp.ac.cn}
	\affiliation{%
		Shanghai Institute of Technical Physics, Chinese Academy of Sciences, Shanghai, 200083, China.}%
	\affiliation{%
		University of Chinese Academy of Sciences, Beijing, 100049, China.}%
	\author{Yun Kau Lau}%
	\email{lau@amss.ac.cn}
	\affiliation{Shanghai Institute of Technical Physics, Chinese Academy of Sciences, Shanghai, 200083, China.}
	\affiliation{
		Institute of Applied Mathematics, Morningside Center of Mathematics, LSSC, Academy of Mathematics and System Science, Chinese Academy of Sciences, 55, Zhongguancun Donglu, Beijing, 100190, China.	}

	\begin{abstract}
		In the detection of gravitational waves in space, during the science phase of the mission, the point ahead angle mechanism (PAAM) serves to steer a laser beam to compensate for the angle generated by the relative motion of the two spacecrafts (SCs) during the approximately 10 seconds of flight time a laser beam will take from one SC to reach a distant SC of three million kilometers away. The common practice for pointing stability control of a laser beam is to  first do a coarse tracking by the PAAM to steer a laser beam to compensate for the relative motion between two SCs, to be followed by a fine pointing stability control. In the present work, by exploiting the near-circular orbit structure of individual SC in the triangular constellation,	the feasibility of inserting an adaptive Kalman filter (AEKF) into the PAAM control loop is investigated. 
		By adopting a colored measurement noise model that closely resembles the prospective on orbit situation, numerical simulation suggests that the dynamic range of the PAAM may be reduced to the level of nano-radians  using the prediction of the pointing head angle (PAA) by the AEKF. 
		This will cut down on the TTL coupling noise and the position noise budget allocated to the PAAM. This in turn reduces the dynamic range of the fine pointing control and leaves room to improve its accuracy, thereby offers the prospect of reduction of the position noise budget allocated to the laser pointing instability as a whole. 
	\end{abstract}
	
	\maketitle
	
	%\tableofcontents
	
	\section{\label{sec1}Introduction}

	In detecting gravitational waves in space, the PAAM serves different functions at different stages of the mission. During the scientific phase of a mission, due to the million kilometers apart between two SCs, a laser beam emanating from one SC will take about ten seconds to reach a distant one. Between this time interval, the distant SC moves, and the PAAM serves to steer a laser beam to compensate for the distance traversed during this interval.	At the same time, it also serves to compensate for the breathing angle generated by solar gravity at the annual level.    
	
	In the present work, we shall address problems related to the laser pointing control in relation to the PAAM stability during the scientific phase. We will confine our attention to the triangular constellation with heliocentric orbit \cite{Taijibrief,lisa}. It is conceivable that the analysis may be applicable to triangular constellation in near earth orbit as well \cite{tianqin}. The conventional practice for pointing stability control of an inter-satellite link is partitioned into two stages. The PAAM performs coarse pointing, to  be followed in the second stage by a more accurate fine pointing control to achieve the accuracy of nano-radian pointing. 
	The micro-radian angular jitter together with the piston noise of PAAM, when coupled with the position noise of the steering mechanism and the phase center offset from the center of mass of a SC, will generate position noise for the gravitational wave measurement.
	
	Given the simple near-circular orbit of each SC in the triangular satellite formation in the detection of gravitational waves in space, it is natural to inquire whether an AEKF based on the orbital dynamics may be incorporated into the PAAM control loop so that the pointing accuracy may be improved and the dynamic range of the PAAM is reduced at the same time. This will cut down on the position noise budget for the PAAM and  mitigate at least partially the TTL noise generated by the PAAM.
	As a preliminary step to explore this line of thought, the aim of the present work is to study this problem more in-depth by means of simulation. The work to be presented in what follows suggests that AEKF does have the potential to play an instrumental role in the PAAM control loop.

	The structure of this paper is organized as follows. Some background materials concerning PAA and AEKF with colored noise are introduced in sections 2 and 3, respectively. Section 4 presents the design of the control loop for the PAAM with the insertion of AEKF. We begin to enter the core of our work in section 5 and set up the AEKF framework for the PAAM.   The system noise model to be employed in the AEKF is worked out in section 6. Section 7 discusses the measurement noise model in the AEKF and the generation of random time-domain noise signals with prescribed PSD used in the simulation. Section 8 presents the simulation results and conducts an in-depth analysis of the filtering results obtained from AEKF. In the final section, some remarks that look to the future of this work are made to conclude our work.

	\section{\label{sec2}PAAM—some basics.}

	In this section, we shall review some basics of the PAAM. This serves to provide background for subsequent discussions and at the same time fix notations and conventions. 
	
	\begin{figure}[hbt!]
		\centering  %居中
		\subfigure[Two-dimensional planar diagram of the PAA.]{   %第一张子图
			\begin{minipage}{0.4\linewidth}
				\centering    %子图居中
				\includegraphics[width=0.7\textwidth,height=0.65\textwidth]{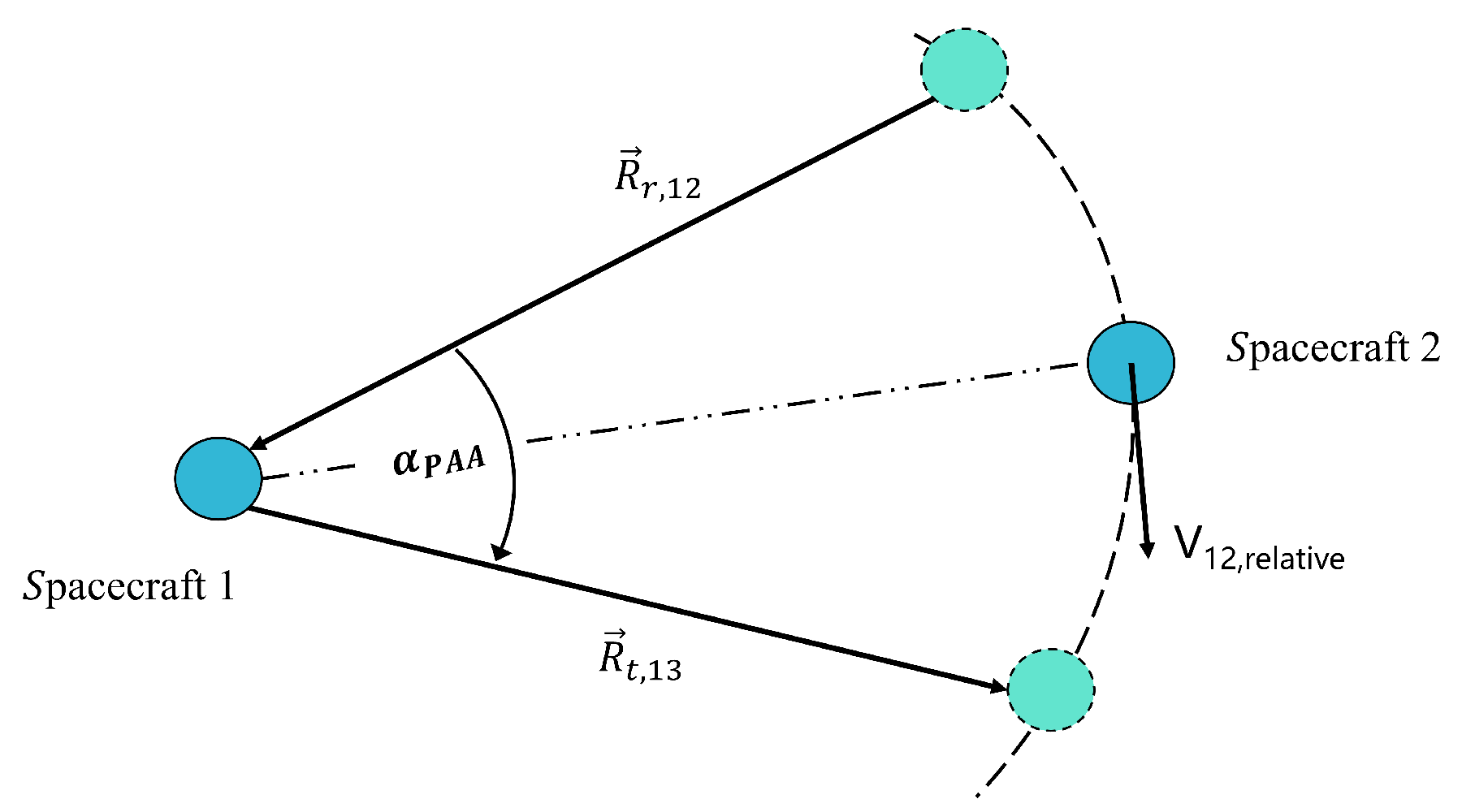}
			\end{minipage}
		}
		\subfigure[Three-dimensional planar diagram of the PAA.]{ %第二张子图
			\begin{minipage}{0.4\linewidth}
				\centering    %子图居中
				\includegraphics[width=0.7\textwidth,height=0.65\textwidth]{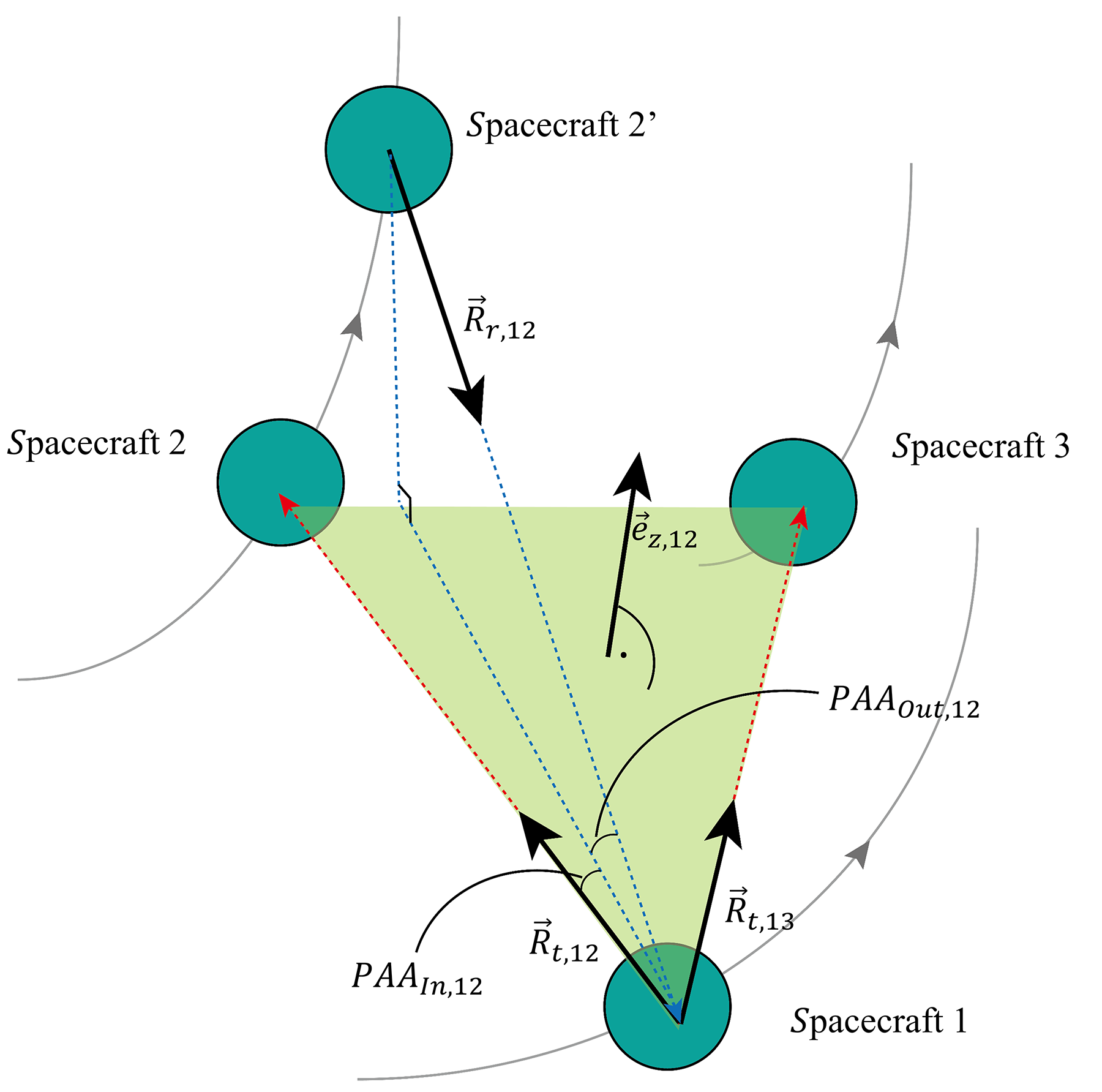}
			\end{minipage}
		}
		\caption{Definition of in-Plane and out-of-Plane PAAs}    %大图名称
		\label{PAASend}    %图片引用标记
	\end{figure}
	PAA is the angle formed between the direction of the transmitted beam between the local SC and the remote SC and the direction of the received beam between the remote SC and the local SC, as shown in Fig.~\ref{PAASend}. The angle varies annually with the orbit of the triangular constellation. Throughout this work, we shall adopt the J2000.0 coordinate system to describe the position and velocity of SCs. The J2000.0 coordinate system is based on the North Celestial Pole and the Vernal Equinox of epoch J2000.0, with the Z axis pointing to the North Celestial Pole, the X axis pointing to the Vernal Equinox, and the X, Y, and Z axes forming a right-handed rectangular coordinate system. The positions and velocities of SCi (i=1,2,3) at a certain epoch may be expressed respectively as $[X_{i}, Y_{i}, Z_{i}]^{T}$, and  $[V_{\tiny {Xi}}, V_{\tiny {Yi}}, V_{\tiny {Zi}}]^{T}$. The relative position and velocity of two SCs are then given by vector addition or subtraction with respect to the J2000.0 reference frame.

	\subsection{Calculations of PAA.}%2级标题
	
	\begin{figure}[hbt!]
		\includegraphics[width=0.8\textwidth,height=0.35\textwidth]{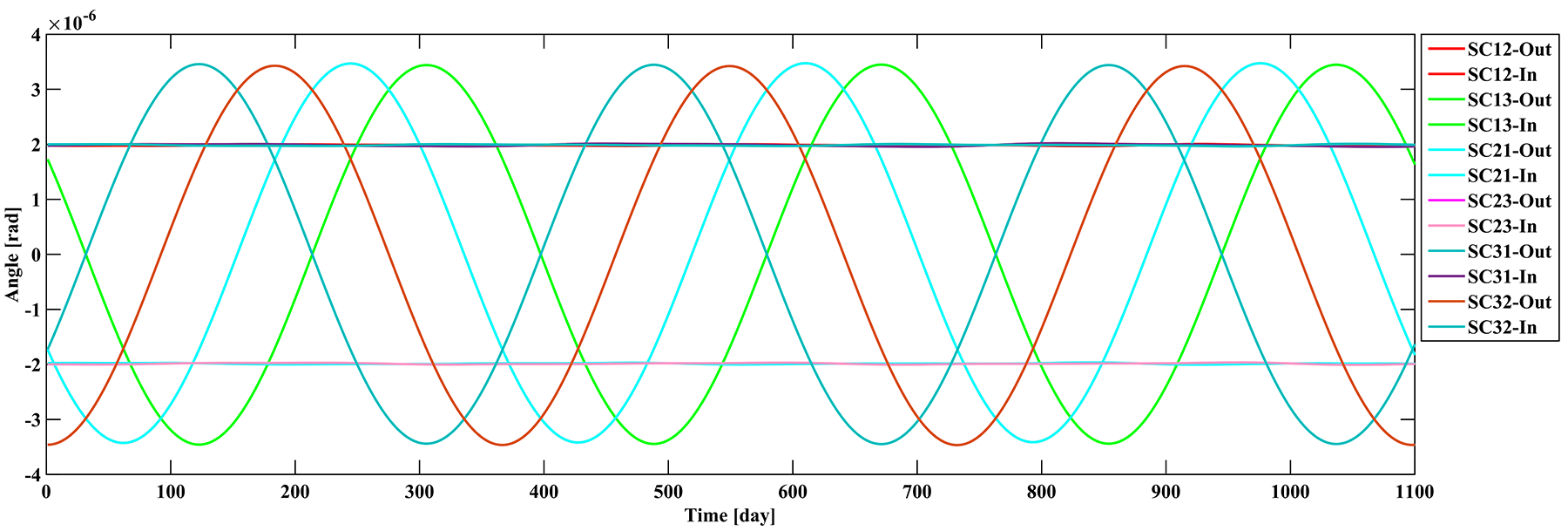}
		\caption{Dynamic range of PAAs.}
		\label{Dynamic Range of PAA}
	\end{figure}
	Fig.~\ref {PAASend} illustrates the in-plane and out-of-plane PAAs. Fig.~\ref{Dynamic Range of PAA} displays the dynamic range of PAAs. Denote by ${\overrightarrow{R}}_{t,ij}$ $i,j=1,2,3$ the beam vector linking a local SC to a remote SC. ${\overrightarrow{R}}_{r,ij}$ is the beam vector linking remote SC to the local SC. The PAA may be calculated to be
	\begin{eqnarray}\label{PAA1} 
		\hbox{PAA} = {\mathit{\arcsin}\left(\bigg| {\frac{{\overrightarrow{R}}_{t,ij}}{|R_{t,ij}|} \times \frac{{\overrightarrow{R}}_{r,ij}}{|R_{r,ij}|}}\bigg|\right)}.
	\end{eqnarray} 
	Take $(i,j)=(1,2)$ as example,   	
	$\Delta t_{r,\ot}$ is the  time taken for a laser beam originating from SC1 to reach SC2 and $ \Delta t_{t,\ot}$ is the time taken for a laser beam from SC2 to reach SC1 upon receiving the laser beam from SC1. We have
	\begin{widetext}
		\begin{eqnarray}\label{PAA3} 
			R_{r,\ot} = \left| {{\overrightarrow{R}}_{\ot} - \Delta t_{r,\ot} \cdot {\overrightarrow{V}}_{\ot}} \right| = c\Delta t_{r,\ot},	
		\end{eqnarray}
		\begin{eqnarray}\label{PAA4} 
			R_{t,\ot} = \left| {{\overrightarrow{R}}_{\ot} + \Delta t_{t,\ot} \cdot {\overrightarrow{V}}_{\ot}} \right| = c\Delta t_{t,\ot},	
		\end{eqnarray}
	\end{widetext}
	Numerical calculations with our orbit integrator enable us to infer that the variation in light time propagation between two SCs is $10\pm 0.013$ seconds in one year, and the resulting change in PAA angle may be neglected \cite{Tang2}. Therefore, we will adopt the approximation that $\Delta t_{r,\ot}=\Delta t_{t,\ot}=\Delta t\approx \,10\;\hbox{seconds}$ in the calculations that follow. 
	
	We shall divide the PAA into two parts in the calculations: in-plane and out-of-plane, as shown in Fig.~\ref{PAASend}.
	
	Define 
	\begin{eqnarray}
		{\overrightarrow{e}}_{x,\ot} = \frac{{\overrightarrow{R}}_{t,\ot}}{|R_{t,\ot}|},\label{e_x}	
	\end{eqnarray}
	\begin{eqnarray}
		{\overrightarrow{e}}_{z,\ot} = \frac{{\overrightarrow{R}}_{t,\ot} \times {\overrightarrow{R}}_{ \small t,\scriptsize{13}}}{\left| {{\overrightarrow{R}}_{t,\ot} 
				\times {\overrightarrow{R}}_{\small t,\scriptsize{13}}} \right|},	
	\end{eqnarray}
	\begin{eqnarray}
		{\overrightarrow{e}}_{y,\ot } = {\overrightarrow{e}}_{{{z,}}\ot } \times {\overrightarrow{e}}_{{{x,}}\ot},\label{e_y}
	\end{eqnarray}
	Which are the unit direction vectors pointing along the X, Z, and Y axes, respectively. 
	\begin{figure}[hbt!]
		\centering
		\includegraphics[width=0.5\textwidth,height=0.25\textwidth]{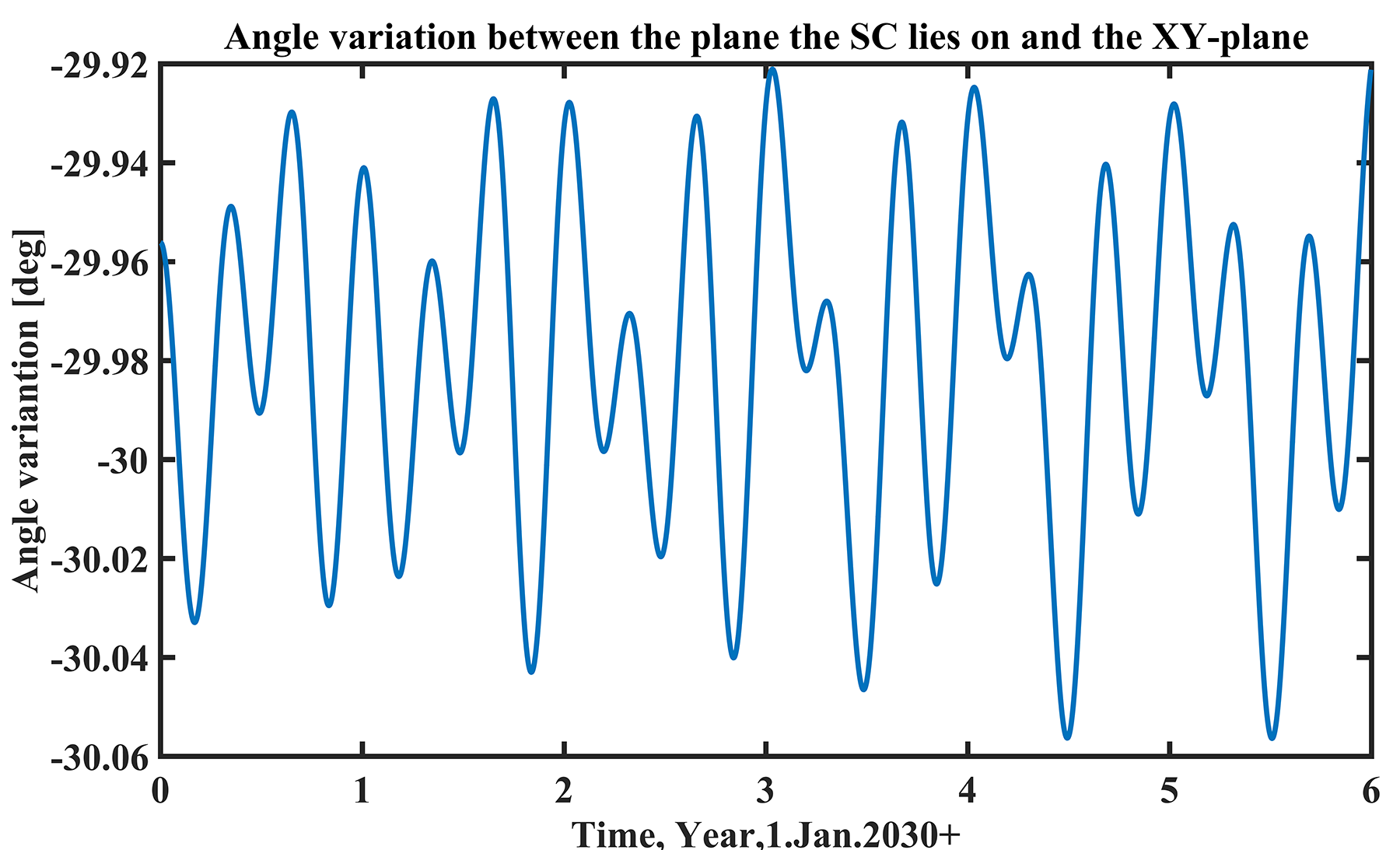}
		\caption{The angle variation between the constellation plane and the Heliocentric Ecliptic Coordinate System of J2000.0.} \label{constellation}
	\end{figure}
	
	Fig.~\ref{constellation} shows the variation of the inclination angle between the triangle constellation plane and the ecliptic plane, with an annual variation range of around 0.08 degrees\cite{Tang2}. 
	This prompts us to approximate ${\overrightarrow{e}}_{{z,}\ot}$ as a constant vector. The error of the in-plane and out-of-plane PAAs due to this approximation is around  0.45 nrad, which is negligible compared to the half-angle of the beam ($\approx$ 1.43 $\mu \hbox{rad}$) \cite{PAAcalculate1}.

	The in-plane PAA is then given by 
	\begin{eqnarray}
		\hbox{PAA}_{\,\hbox{\footnotesize  in,}\scriptsize{12}}={\sin}^{- 1}\left( {\frac{ {{\overrightarrow{R}}_{r,\ot} \times {\overrightarrow{e}}_{\small z,\ot}} }{\left| {{\overrightarrow{R}}_{r,\ot} \times {\overrightarrow{e}}_{\small z,\ot}} \right|} \cdot {\overrightarrow{e}}_{\small x,\ot }} \right)\approx  {\frac{ {{\overrightarrow{R}}_{r,\ot} \times {\overrightarrow{e}}_{\small z,\ot}} }{\left| {{\overrightarrow{R}}_{r,\ot} \times {\overrightarrow{e}}_{\small z,\ot}} \right|} \cdot {\overrightarrow{e}}_{\small x,\ot}} \label{PAA_IN}.	
	\end{eqnarray}
	The out-of-plane PAA may be  expressed as
	\begin{eqnarray}
		\hbox{PAA}_{\,\hbox{\footnotesize out,\scriptsize{12}}} = {\mathit{\sin}^{- 1}\left( \frac{{\overrightarrow{e}}_{\small z,\ot} \cdot {\overrightarrow{R}}_{r,\ot}}{|\overrightarrow{R}_{r,\ot}|} \right)\approx \frac{{\overrightarrow{e}}_{z,\ot} \cdot {\overrightarrow{R}}_{r,\ot}}{|\overrightarrow{R}_{r,\ot}|} }\label{PAA_OUT}.	
	\end{eqnarray}
	In Eq.~(\ref{PAA_IN}) and Eq.~(\ref{PAA_OUT}), we have used the approximation that ${\sin}x\approx x$, given that both  $\hbox{PAA}_{\,\hbox{\tiny out,\scriptsize{12}}}$ and $ \hbox{PAA}_{\,\hbox{\tiny out,\scriptsize{12}}}$ are of the order of $\mu \hbox{rad}$.

	In the next step, we shall express  the relative position and velocity vector between SCs in terms of the coordinate components in the J2000.0 frame. Let 
	\begin{align}
		\vec{R}_{12}=[x_{12},y_{12},z_{12}]=[ X_{1}- X_{2}, Y_{1}- Y_{2}, Z_{1}- Z_{2}],
	\end{align}	
	represent the  laser link directed from SC1 to SC2. Likewise, 
	\begin{align}
		\vec{V}_{12}=[V_{\tiny x_{12}},V_{\tiny y_{12}},V_{\tiny z_{12}}]=[ V_{\tiny {X1}}- V_{\tiny {X2}}, V_{\tiny {Y1}}- V_{\tiny {Y2}}, V_{\tiny {Z1}}- V_{\tiny {Z2}}],
	\end{align} 
	is the  relative velocity vector from SC1 to SC2.
	$[X_{i},Y_{i},Z_{i}]$ is the current position information of SCi with $i=1,2,3$, and $[V_{\tiny {Xi}},V_{\tiny {Yi}},V_{\tiny {Zi}}]$ is the current velocity information of SCi. The beam vector $\vec{R}_{t,\ot}$ and beam vector $\vec{R}_{r,\ot}$  linking SC1 to SC2 may be expressed as 
	\begin{align}
		\vec{R}_{t,\ot}=[{x_{12}+V_{\tiny x_{12}}\Delta t},{y_{12}+V_{\tiny y_{12}}\Delta t},{z_{12}+V_{\tiny z_{12}}\Delta t}],
	\end{align}	
	\begin{align}
		\vec{R}_{r,\ot}=[{x_{12}-V_{\tiny x_{12}}\Delta t},{y_{12}-V_{\tiny y_{12}}\Delta t},{z_{12}-V_{\tiny z_{12}}\Delta t}]\label{Rr12}.
	\end{align}
	The beam vector $\vec{R}_{\small t,\scriptsize{13}}$ linking SC1 to SC3 may be expressed as 
	\begin{align}
		\vec{R}_{ \small t,\scriptsize{13}}=[{x_{13}+V_{\tiny x_{13}}\Delta t},{y_{13}+V_{\tiny y_{13}}\Delta t},{z_{13}+V_{\tiny z_{13}}\Delta t}].
	\end{align}
	According to Eq.~(\ref{e_x})-Eq.~(\ref{e_y}), the X-axis directional vector of the SC1 may be expressed as
	\begin{eqnarray}
		{\overrightarrow{e}}_{x,\ot} = \frac{[{x_{\ot }+Vx_{\ot }\Delta t},{y_{12}+V_{\tiny y_{12}}\Delta t},{z_{12}+V_{\tiny z_{12}}\Delta t}]}{\left|{\overrightarrow{R}}_{t,\ot}\right|}\label{ex}.	
	\end{eqnarray}
	Write  ${\overrightarrow{e}}_{{{z,}}\ot}$ in terms of the coordinate components of J2000.0 as 
	\begin{eqnarray}
		&&	{\overrightarrow{e}}_{{{z,}}\ot}
		= \begin{bmatrix}
			{{e_z}_{x}},
			{{e}_{zy}}, 
			{{e}_{zz}}
		\end{bmatrix}	
		\label{ez}.
	\end{eqnarray}
	The Y-axis direction vector of SC1 may also be be expressed in terms of the coordinate components of J2000.0 as
	\begin{eqnarray}
		{\overrightarrow{e}}_{{{y,}}\ot} 
		=\frac{1}{\left|{\overrightarrow{R}}_{t,\ot}\right|}
		\begin{bmatrix}
			({z_{12}+V_{\tiny z_{12}}\Delta t}){e}_{zy}-({z_{12}+V_{\tiny z_{12}}\Delta t}){e}_{zz}\\
			({x_{12}+V_{\tiny x_{12}}\Delta t}){e}_{zz}-({z_{12}+V_{\tiny z_{12}}\Delta t}){e}_{zx}\\
			({y_{12}+V_{\tiny y_{12}}\Delta t}){e}_{zx}-({x_{12}+V_{\tiny x_{12}}\Delta t}){e}_{zy}
		\end{bmatrix}^T.	
	\end{eqnarray}
	
	Then the expression of in-plane PAA is given by 
	\begin{eqnarray}
		\hbox{PAA}_{\hbox{\tiny in,\scriptsize{12}}}
		={\frac{ {{\overrightarrow{R}}_{r,\ot} \times {\overrightarrow{e}}_{z,\ot}} }{\left| {{\overrightarrow{R}}_{r,\ot} \times {\overrightarrow{e}}_{z,\ot}} \right|} \cdot {\overrightarrow{e}}_{x,\ot}}
		=\frac{{\overrightarrow{R}}_{r,\ot} \times {\overrightarrow{e}}_{z,\ot}\cdot {\overrightarrow{e}}_{x,\ot}}
		{\left|\overrightarrow{R}_{r,\ot}\right|\left|\overrightarrow{R}_{t,\ot}\right|}.
	\end{eqnarray}
	with 
	\begin{eqnarray}
		&&{\overrightarrow{R}}_{r,\ot} \times {\overrightarrow{e}}_{z,\ot}\cdot {\overrightarrow{e}}_{x,\ot}\nonumber\\
		&=&{(({y_{12}-V_{\tiny y_{12}}\Delta t}){e}_{zz}-({z_{12}-V_{\tiny z_{12}}\Delta t}){e}_{zy})({x_{12}+V_{\tiny x_{12}}\Delta t})}\nonumber\\
		&&+(({z_{12}-V_{\tiny z_{12}}\Delta t}){e}_{zx}-({x_{12}-V_{\tiny x_{12}}\Delta t}){e}_{zz})({y_{12}+V_{\tiny y_{12}}\Delta t})\nonumber\\
		&&+(({x_{12}-V_{\tiny x_{12}}\Delta t}){e}_{zy}-({y_{12}-V_{\tiny y_{12}}\Delta t}){e}_{zx})({z_{12}+V_{\tiny z_{12}}\Delta t}).
		\label{E_1}	
	\end{eqnarray}
	The out-of-plane PAA is expressed as
	\begin{eqnarray}
		\hbox{PAA}_{\hbox{\tiny out,\scriptsize{12}}} =\frac{{\overrightarrow{e}}_{z,\ot } \cdot {\overrightarrow{R}}_{r,\ot}}{|\overrightarrow{R}_{r,\ot}|}.
	\end{eqnarray}
	with 
	\begin{eqnarray}
		&& {\overrightarrow{e}}_{z,\ot } \cdot {\overrightarrow{R}}_{r,\ot}\nonumber\\
		&=& {e}_{zx}({x_{12}-V_{\tiny x_{12}}\Delta t})+{e}_{zy}({y_{12}-V_{\tiny y_{12}}\Delta t})+{e}_{zz}({z_{12}-V_{\tiny z_{12}}\Delta t}).
		\label{E_2}
	\end{eqnarray}
	
	\section{\label{sec3}Adaptive extended Kalman Filter}%

	In this section, we shall introduce the AEKF method, on the basis of which we develop a new control algorithm for the PAAM. The material presented in this section is not new \cite{wangyan1}, but to render the present article self-contained, we will describe the AEKF framework and then apply it to the PAAM  in the next section.

	Based on the standard EKF\cite{EKF_origin,wangyan1}, we can treat the measurement noise as a state quantity and include it in the state equation \cite{colorrangingnoise,colorrangingnoise1}. The state equation of the Kalman filter, which includes the extended system noise and the measurement noise, is obtained as Eq.~(\ref{statement}).
	\begin{eqnarray}
		\begin{bmatrix}
			X_{k + 1} \\
			v_{k + 1} \\
		\end{bmatrix} = \begin{bmatrix}
			\phi_{k,k - 1} & 0\\
			0 & \Psi_{k ,k-1} \\
		\end{bmatrix} \begin{bmatrix}
			X_{k} \\
			v_{k} \\
		\end{bmatrix} + \begin{bmatrix}
			\Gamma_{k + 1,k}  & 0 \\
			0 & I\\
		\end{bmatrix}\begin{bmatrix}
			W_{k}  \\
			\xi_{k} \\
		\end{bmatrix}\label{statement}.
	\end{eqnarray}
	Where $\Gamma_{k + 1,k}$ is an identity matrix. Based on the original measurement equation(\ref{measurement}), make
	\begin{eqnarray}
		Z_{k}^{*} = Z_{k + 1} - \Psi_{k ,k-1} \times Z_{k + 1}.	
	\end{eqnarray}
	It is obtained by substituting the expression $Z_{ k }$
	\begin{eqnarray}
		Z_{k + 1} - \Psi_{k ,k-1}Z_{k} = \left( {H_{k + 1}\phi_{k + 1,k} - \Psi_{k ,k-1}H_{k}} \right)X_{k} + H_{k + 1}\Gamma_{k}W_{k} + \xi_{k}.	
	\end{eqnarray}
	We define $( {H_{k + 1}\phi_{k + 1,k} - \Psi_{k ,k-1}H_{k}}) $ as $H_{k}^{*}$. $( {H_{k + 1}\Gamma_{k}W_{k} + \xi_{k}} ) $ is defined as $ V_{k}^{*} $. Thus, a new measurement equation is established.
	\begin{eqnarray}
		Z_{k}^{*} = H_{k}^{*}X_{k} + V_{k}^{*},	
	\end{eqnarray}
	Where $ V_{k}^{*} $ is zero mean white noise and its variance is
	\begin{eqnarray}
		R_{k}^{*} = H_{k + 1}\Gamma_{k}Q_{k}\Gamma_{k}^{T}H_{k + 1}^{T} + R_{k}.	
	\end{eqnarray}
	In this case, the covariance matrix of the measurement noise and the process noise is
	\begin{eqnarray}
		S_{k} = Q_{k}\Gamma_{k}^{T}H_{k + 1}^{T}+ \xi_{k}\label{start}.	
	\end{eqnarray}
	The expression for the posterior estimation equation is
	\begin{eqnarray}
		\hat X_{k + 1} = \phi_{k + 1,k}\hat X_{k} + {\bar K}_{k + 1}( Z_{k + 1} - \Psi_{k ,k-1}Z_{k} - H_{k}^{*}{\hat X_{k}}).	
	\end{eqnarray}
	The expression of new Kalman gain is
	\begin{eqnarray}
		\bar {K}_{k + 1} = \left( {\phi_{k + 1,k}P_{k}H_{k}^{*\ T} + \Gamma_{k}S_{k}} \right)\left( {H_{k}^{*}P_{k}H_{k}^{*\ T} + R_{k}^{*}} \right)^{- 1}.	
	\end{eqnarray}
	The process noise covariance matrix is
	\begin{eqnarray}
		p_{k + 1}^{*} = \phi_{k + 1,k}P_{k}^{*}\phi_{k + 1,k}^{T} + \Gamma_{k}Q_{k}\Gamma_{k}^{T} - {\bar{K}}_{k + 1}\left( {H_{k}^{*}P_{k}^{*}\phi_{k + 1,k}^{T} + S_{k}^{T}\Gamma_{k}^{T}} \right)\label{end}.	
	\end{eqnarray}
	The formula of EKF algorithm subject to colored measurement noise is given by Eq. (\ref{start})-Eq.~(\ref{end}).
	
	\section{\label{sec4}PAAM control loop design }

\begin{figure}[hbt!]
		\centering
		\includegraphics[width=0.7\textwidth,height=0.25\textwidth]{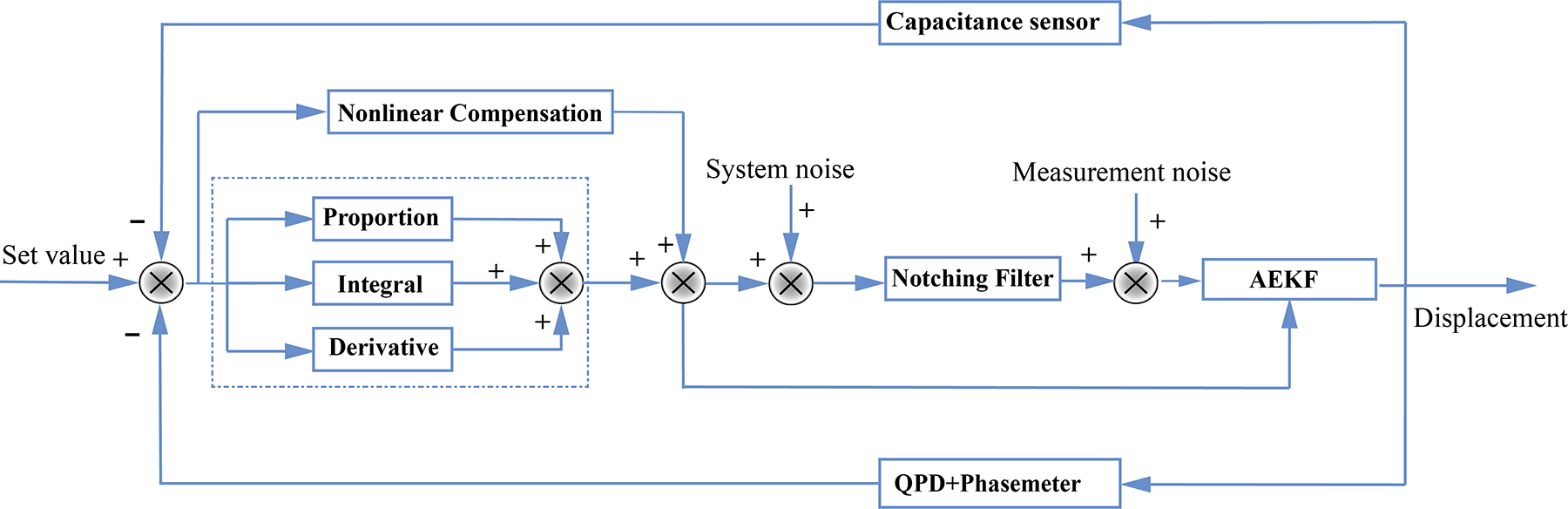}
		\caption{PAAM control framework.} \label{controlloop}
	\end{figure}

	In a LISA-type mission,  unlike the EKF considered before for clock synchronization purposes in the pre-TDI data post-processing \cite{wangyan2}, the AEKF for PAAM will be carried out on orbit. 
	In our design of the PAAM control, the FPGA control chip receives the feedback signal from the four-quadrant detector demodulated by the phasemeter and the capacitance sensors of PAAM. It combines the information of the SC orbit integrator for feedforward control. Fig \ref{controlloop} shows the block diagram design of the control system of the PAAM. In this control system, a theoretical model is established by AEKF and SC orbital integrator to control the PAAM. The output value controlled by the PID is weighted with the system noise as the state input to AEKF, and the measurement noise is added to simulate the actual situation on the SC. The whole control loop is closed-loop controlled by a phasemeter demodulated four-quadrant detector and capacitance sensors. Taking into account the creep and hysteresis of piezoelectric ceramics in the PAAM, nonlinear compensation, and notch filters are added to the control loop to improve the stability of the whole control loop.
	
	The sampling frequency is chosen to be one per day, for the reason that the PAA varies slowly with the orbit. Further, in every manoeuvre of the PAAM, the piston noise couples with spacecraft attitude jitter and possibly other noise sources and generate position noise for gravitational wave measurement. A low sampling frequency will minimise  the disturbance in the laser interferometry measurement. 
	
	Unlike the EKF considered before for clock synchronization purposes in the pre-TDI data post-processing \cite{wangyan3}, the AEKF for PAAM will be carried out on orbit, and input of measurement data from the precision orbit determination of SCs
	is required. As the distance between the Earth and the triangle constellation is very similar to that between Mars and the Earth, the experience for the precision orbit determination of the Martian mission Tianwen I \cite{yangpeng} will serve as a useful reference guide. 
	As the sampling frequency is one data per day, a straightforward way will be for the ground tracking station to upload the orbit data once per day. The delay for communication between the ground tracking station and SC (around half an hour) will be negligible due to the adiabatic, very slow change in the orbit of the triangle constellation. However, from the viewpoint of autonomous navigation and from the experience of the Tianwen I  mission, it is feasible to use orbit prediction data for a month in the measurement model of the AEKF, provided the required precision in orbit determination on the position and velocity of SC orbit are respectively within the margins of 20km and 2cm/s \cite{li2021orbit}.  We can also update the orbit determination error by ground tracking during data transmission between the spacecraft and the ground station.

	\section{\label{sec5}Adaptive extended Kalman Filter Model for PAAM}%
	First, we define a 18-dimensional column state vector, which includes the position and velocity information of three SCs.
	\begin{eqnarray}
		X_{k} = [X_1,Y_1,Z_1,V_{X1},V_{Y1},V_{Z1},X_2,Y_2,Z_2,V_{X2},V_{Y2},V_{Z2},X_3,Y_3,Z_3,V_{X3},V_{Y3},V_{Z3}]^T.
	\end{eqnarray}
	The dynamics  of a single SC is described by Keplerian equation for planetary motion  given by 
	\begin{eqnarray}
		\ddot{\vec{X_{k}}} = \sum_{planet}GM_{planet}\frac{\vec{X_{k}}-\bm{R}_{planet}}{\mid \vec{X_{k}}-\bm{R}_{planet}\mid^{3}},
	\end{eqnarray}
	Where $\vec{X_{k}}$ is the position of one SC, $M_{planet}$, $\bm{R}_{planet}$ are the mass and the coordinates of the ith celestial body (the Sun and the planets) in the solar system, ${\vec{X_{k}}-\bm{R}_{planet}}$ is a vector pointing from that SC to the $i$ th celestial body. In designing the orbit used in our study, we considered only the gravitational forces given by the Sun and the major planets of the solar system when constructing the equations of motion for SC. The major planets include Mercury, Venus, Earth+Moon, Mars, Jupiter, Saturn, Uranus, and Neptune \cite{Tang2}. The dynamic equation can be written in a different form.
	\begin{eqnarray}
		\frac{d}{dt}
		\begin{bmatrix}
			\vec{X_{k}}\\
			\vec{v_{k}}
		\end{bmatrix}
		&=f\left(\vec{X_{k}},\vec{v_{k}}\right)
		&=	\begin{bmatrix}
			\vec{v_{k}}\\
			\sum_{planet}GM_{planet}\,\frac{\vec{X_{k}}-\bm{R}_{planet}}{\mid \vec{X_{k}}-\bm{R}_{planet}\mid^{3}}
		\end{bmatrix}.
	\end{eqnarray}
	We define $\alpha =\left(\vec{X_{k}},\vec{v_{k}}\right)^T$, then we have
	\begin{eqnarray}
		\phi=\frac{\partial f}{\partial \alpha}=
		\begin{bmatrix}
			O_3 & I_3\\
			A   &  O
		\end{bmatrix}.
	\end{eqnarray}
	Here, $O_3$ is the $3\times3$ zero matrix, $I_3$ is the $3\times3$ identity matrix, and the expression of the $3\times3$ matrix A is given as:
	\begin{eqnarray}
		A	&=&-\sum_{planet}\frac{GM_{planet}}{\mid \vec{X_{k}}-\bm{R}_{planet}\mid^{3}}\mathbf{I}_3\nonumber\\
		&&  + \sum_{planet}\frac{3GM_{planet}}{\mid \vec{X_{k}}-\bm{R}_{planet}\mid^{5}}\,(\vec{X_{k}}-\bm{R}_{planet})(\vec{X_{k}}-\bm{R}_{planet})^T.
	\end{eqnarray}
	For the entire system, the dynamic matrix $\phi=\frac{\partial f}{\partial x}$ is $18\,\times\,18$. We omit its explicit expression here, as it can be obtained in a straightforward way from the above formulae.
	In our work that follows, we simulate the PAA data of 3 years. Since the PAA changes very slowly with one year periodicity, we set the sampling frequency of the  AEKF to 1 day. A longer time span up to a few days is also feasible. In the initial design process, we have considered the influence of SCs displacement, velocity, and acceleration on PAA, but in our subsequent simulation, we find that the relative acceleration of the SC is irrelevant to the PAA calculations. 
	
	Next, we shall present the two-dimensional measurement equation. The measurement equation which links up the positions and velocities of the three SCs and the in-plane and out-of-plane PAAs and may be written as 
	\begin{eqnarray}
		Z_{k} = h_{k}\left( {X_{k},v_{k}} \right) = \left\lbrack \hbox{PAA}_{\hbox{\tiny out,\scriptsize{12}}},\hbox{PAA}_{\hbox{\tiny in,\scriptsize{12}}}~\right\rbrack~,
	\end{eqnarray}  
	where $\hbox{PAA}_{\hbox{\tiny in,\scriptsize{12}}}$ and $ \hbox{PAA}_{\hbox{\tiny out,\scriptsize{12}}} $ are respectively the in-plane PAA and the out-of-plane PAA. $v_{k}$ is the measurement nosie. $H_{k}$ is a 2×18-dimensional observation matrix. We omit the explicit of the 36 components in $H_{k}$. The element $H_{k}[i,j]$ in the matrix $H_{k}$ may be expressed as:
	\begin{eqnarray}
		H_{k}[i,j]=\frac{\partial Z_{k}[i]}{\partial X_{k}[j]}.
	\end{eqnarray}
	As an example, the [1,1] and [2,1] components of $H_{k}$, with the step index $k$ omitted, may be given as follows.
	\begin{eqnarray}
		H[1,1]=\frac{\partial\, \hbox{PAA}_{\hbox{\tiny out},\ot}}{\partial X_1}
		=\frac{{e}_{zx}|\overrightarrow{R}_{r,\ot}|^2-({\overrightarrow{e}}_{z,\ot } \cdot {\overrightarrow{R}}_{r,\ot})(x_{12}-V_{\tiny x_{12}}\Delta t)}{|\overrightarrow{R}_{r,\ot}|^3}.
	\end{eqnarray}
	${\overrightarrow{e}}_{z,\ot } \cdot {\overrightarrow{R}}_{r,\ot}$ is given in Eq.~(\ref{E_2}).
	\begin{eqnarray}
		H[2,1]=\frac{\partial\, \hbox{PAA}_{\hbox{\tiny in,\scriptsize{12}}}}{\partial X_1}	
		=\frac{\frac{\partial {\overrightarrow{R}}_{r,\ot} \times {\overrightarrow{e}}_{z,\ot}\cdot {\overrightarrow{e}}_{x,\ot}\nonumber}{\partial X_1}-2x_{12}({\overrightarrow{R}}_{r,\ot} \times {\overrightarrow{e}}_{z,\ot}\cdot {\overrightarrow{e}}_{x,\ot})}{ {|\overrightarrow{R}_{r,\ot}|}{|\overrightarrow{R}_{t,\ot}|}}.
	\end{eqnarray}
	with 
	\begin{eqnarray}
		&&\frac{\partial {\overrightarrow{R}}_{r,\ot} \times {\overrightarrow{e}}_{z,\ot}\cdot {\overrightarrow{e}}_{x,\ot}\nonumber}{\partial X_1}\nonumber\\
		&=&{(({y_{12}-V_{\tiny y_{12}}\Delta t}){e}_{zz}-({z_{12}-V_{\tiny z_{12}}\Delta 	t}){e}_{zy})}\nonumber\\
		&&-{e}_{zz}({y_{12}+V_{\tiny y_{12}}\Delta t})+{e}_{zy}({z_{12}+V_{\tiny z_{12}}\Delta t}),	
	\end{eqnarray}
	with ${\overrightarrow{R}}_{r,\ot} \times {\overrightarrow{e}}_{z,\ot}\cdot {\overrightarrow{e}}_{x,\ot}$ being given in Eq.~(\ref{E_1}).
	
	In standard practice, the coefficient transfer matrix $\Psi_{k,k-1}$ of colored measurement noise is determined by the ARMA model. In the present context, the definition of $\Psi_{k,k-1}$ is relatively simple and can be considered as a special case of the ARMA model in which the autoregressive parameter is 1, and the moving average parameter is 0 \cite{ARMR}. We have also tried to use more sophisticated  ARMA model to estimate the value of the $\Psi_{k,k-1}$ matrix, but the results are not as good as the methods used here. As shown in Eq.~(\ref{statement}), we only consider the colored noise component in the measurement noise and did not include white noise. Therefore, the $\xi_{k}$ term can be directly ignored. As shown in Eq.~(\ref{Psi}), we approximately take the ratio of the measurement noise amplitudes at the previous moment and the current moment as input to the coefficient transfer matrix $\Psi_{k,k-1}$ of the colored measurement noise. The simple choice here is likely due to a very slow variation of the colored measurement noise in the time domain at an annual level. This colored measurement noise is a linear superposition of all colored noise, which we will discuss in detail in a moment. 
	\begin{eqnarray}
		{\Psi_{k,k - 1}} =v_{k}/ v_{k - 1}\label{Psi}.	
	\end{eqnarray}
	
	In a standard adaptive Kalman filter, the size of the $Q_{k}$ and $R_{k}$ matrix is automatically adjusted by observing the prediction error and its mean square error matrix and introducing the fading factor to obtain a good estimation state \cite{adaptiveEKF}. The AEKF designed in this paper is slightly different from the traditional AEKF. In the AEKF designed by us, the covariance matrix of measurement noise $R_k$ is updated in real-time according to the magnitude of measurement noise, while the covariance matrix of system noise $Q_k$ is updated every month or so, based on the accuracy of orbit prediction. It can also be updated by ground tracking during data transmission between the spacecraft and the ground station.
	
	\begin{figure}[hbt!]
		\centering
		\includegraphics[width=0.8\textwidth,height=0.45\textwidth]{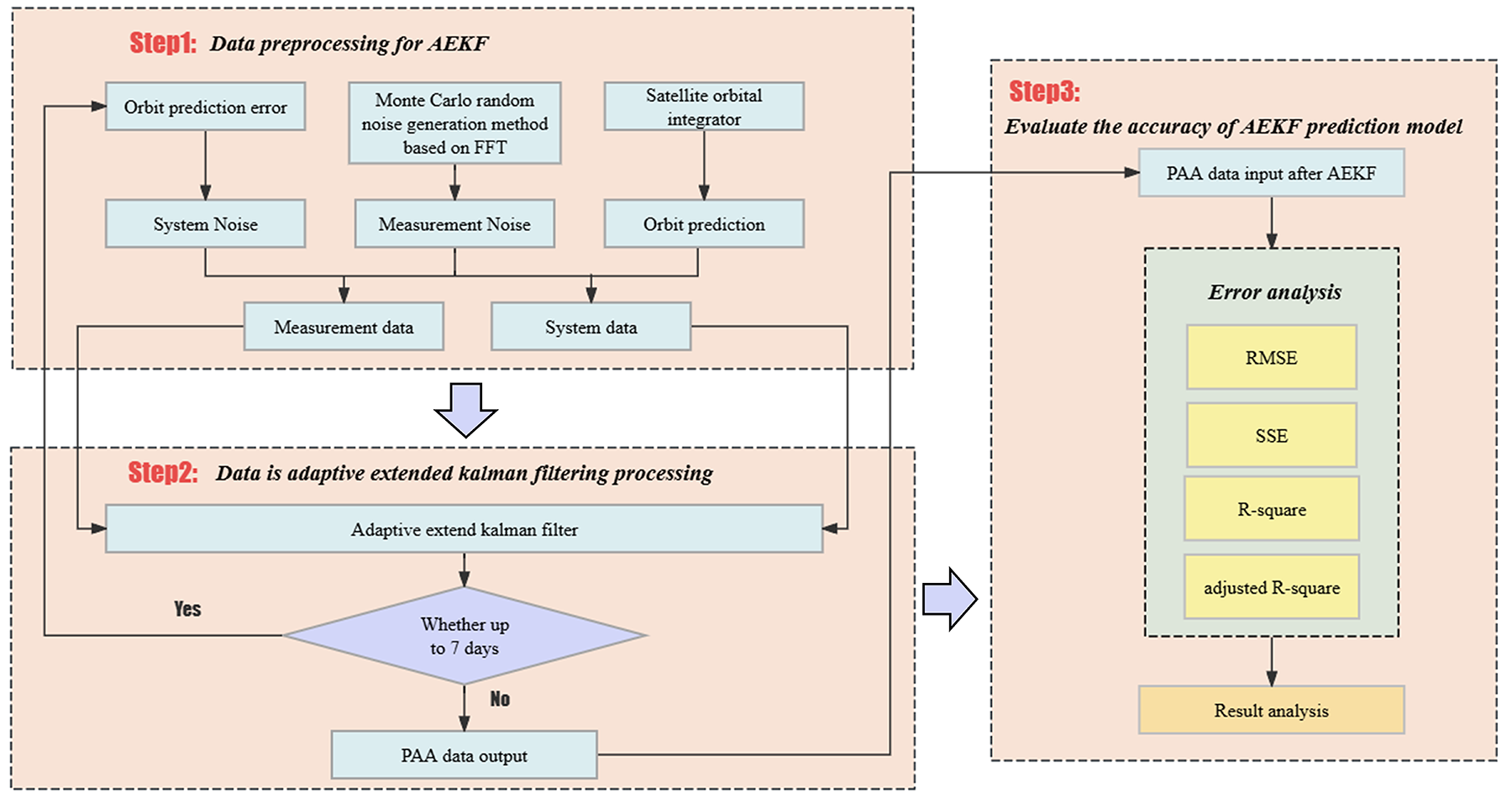}
		\caption{PAA calculation flow chart.} \label{PAA calculation}
	\end{figure}
	
	In this study, the flowchart for PAA adjustment computation is illustrated in Fig.~\ref{PAA calculation}. During the first step of data preprocessing, we rely on orbit prediction to acquire the SC's position and velocity information. The error in orbit prediction within this system is linearly superimposed onto the orbit prediction data as system noise to form the system data. The orbit prediction data, linearly superimposed with noise generated by the FFT-based Monte Carlo random noise generation method, constitutes the measurement data input into the AEKF. The system data and measurement data are generated using different orbital prediction data, each based on different points of the orbital determination error as input. At the same time, the orbital determination error can also be obtained without orbital prediction. During data transmission between the spacecraft and the ground station, ground tracking can also be used to update it. In the second step, during the AEKF process, the orbit prediction information needs to be updated every month. In the third step, we conduct an error analysis on the data filtered by AEKF, using four independent methods: SSE (Sum of Squared Errors), RMSE (Root Mean Squared Error), R-square, and adjusted R-square for error fitting.
	Finally, the results are analyzed and discussed.
	
	\section{\label{sec6} System Noise Model}
	To work out a system model for AEKF, consider
	\begin{eqnarray}
		\left\{\begin{array}{l}{\delta X} = [\delta x,\delta y,\delta z]^T,
			\\\delta V= [\delta V_x,\delta V_y,\delta V_z]^T.
		\end{array}\right.
		\label{error}
	\end{eqnarray}	
	where ${\delta X}$ and $\delta V$ are the errors in respectively the position and velocity in the orbit prediction. We take the SC's position and velocity data generated by the orbit integrator as input into the AEKF while incorporating the orbit prediction error in the form of system noise.
	Accordingly, we express the SC position coordinates $[X_{i}+\delta x, Y_{i}+\delta y, Z_{i}+\delta z]^{T}$ and velocity coordinates $[V_{\tiny {Xi}}+\delta V_x, V_{\tiny {Yi}}+\delta V_y, V_{\tiny {Zi}}+\delta V_z]^{T}$,  Eq.~(\ref{PAA1}) - Eq.~(\ref{PAA_OUT}) are then employed to calculate the PAA with orbital prediction errors taken into account. The position coordinates$[X_{i}, Y_{i}, Z_{i}]^{T}$ and velocity coordinates $[ V_{\tiny {Xi}}, V_{\tiny {Yi}}, V_{\tiny {Zi}}]^{T}$ of the SC, free of orbital prediction errors, will be used to compute the true value of PAA by substituting them into Eq.~(\ref{PAA1})-Eq.~(\ref{PAA_OUT}). The difference between these two values represents the PAA  error generated by the orbital prediction. This error is input as the source of system noise for the AEKF. We then transform this orbit prediction error into both in-plane and out-of-plane PAA errors, resulting in an error margin of approximately 0.2 $\hbox{nrad}$. The system noise is linearly superimposed onto the genuine PAA value provided by the orbital integrator in the form of white noise. Additionally, the covariance matrix $Q_{k}$ is configured in accordance with the system noise.

	\section{\label{sec7}Measurement Noise Model}
	This section primarily elaborates on the PAA noise model, serving as the foundation for quantifying measurement noise in our subsequent simulations. During the simulation process, position and velocity of SCs are generated by an the orbit integrator to mimic orbit prediction. We consider this to be the true value  and linearly superimpose it with the measurement noise data, which serves as the input of the measurement value for AEKF. To better simulate the actual conditions on orbit, we adopt an FFT-based Monte Carlo method to generate random time-domain noise signals with specified PSD. Specifically, the colored noise generated by this method is produced using the LISA noise PSD in the time domain with the help of the LTPDA toolkit \cite{ltpda}. At the same time,  the covariance matrix $R_k$, which characterizes the measurement noise within the AEKF, is dynamically computed from the generated noise and updated in real-time \cite{colorrangingnoise2} as
	\begin{eqnarray} 
		R_{k} = \hbox{cov}[v_{k},v_{k}]\label{R_update}.
	\end{eqnarray}
	The noise model accounts for various sources, and we add a total of $10\times 10^{-12}\frac{m}{\sqrt{Hz}}$ measurement noise, taking into account mainly piston noise of PAAM, directional jitter noise, SC attitude jitter noise, readout noise among others. In AEKF, since the measurement noise is much larger than the system noise, the size of its prediction error mainly depends on the size of the system noise. Therefore, increasing the measurement noise does not significantly affect the prediction results. Then, we use the LTPDA toolbox to generate time-domain noise based on the noise power spectral density function. The sum of the measurement noise can be expressed as Eq.~(\ref{noise}), and f is the frequency.
	\begin{eqnarray}  
		\overset{\sim}{x}_{\text{noise}}(f) = 10 \times 10^{-12} \sqrt{1 + \left( \frac{2.8 \, \text{mHz}}{f} \right)^{4}} \, \frac{\text{m}}{\sqrt{\text{Hz}}}.  
		\label{noise}  
	\end{eqnarray}
	\begin{figure}[ht!]
		\centering
		\includegraphics[width=0.4\textwidth,height=0.4\textwidth]{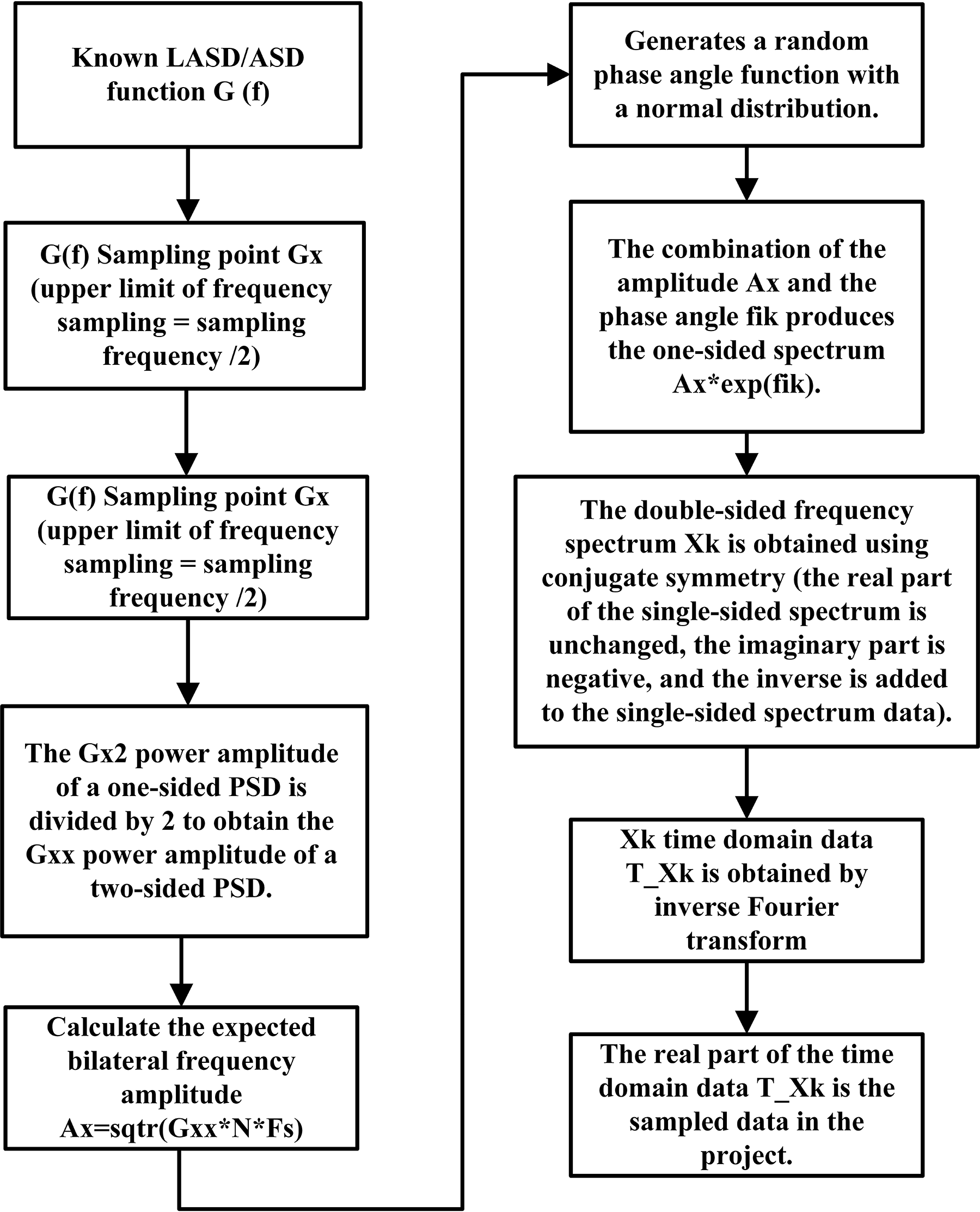}
		\caption{Flow chart of time domain noise acquisition.} \label{ltpda}
	\end{figure}
	
\begin{figure}[htbp]
		\centering
		\includegraphics[width=0.5\textwidth,height=0.4\textwidth]{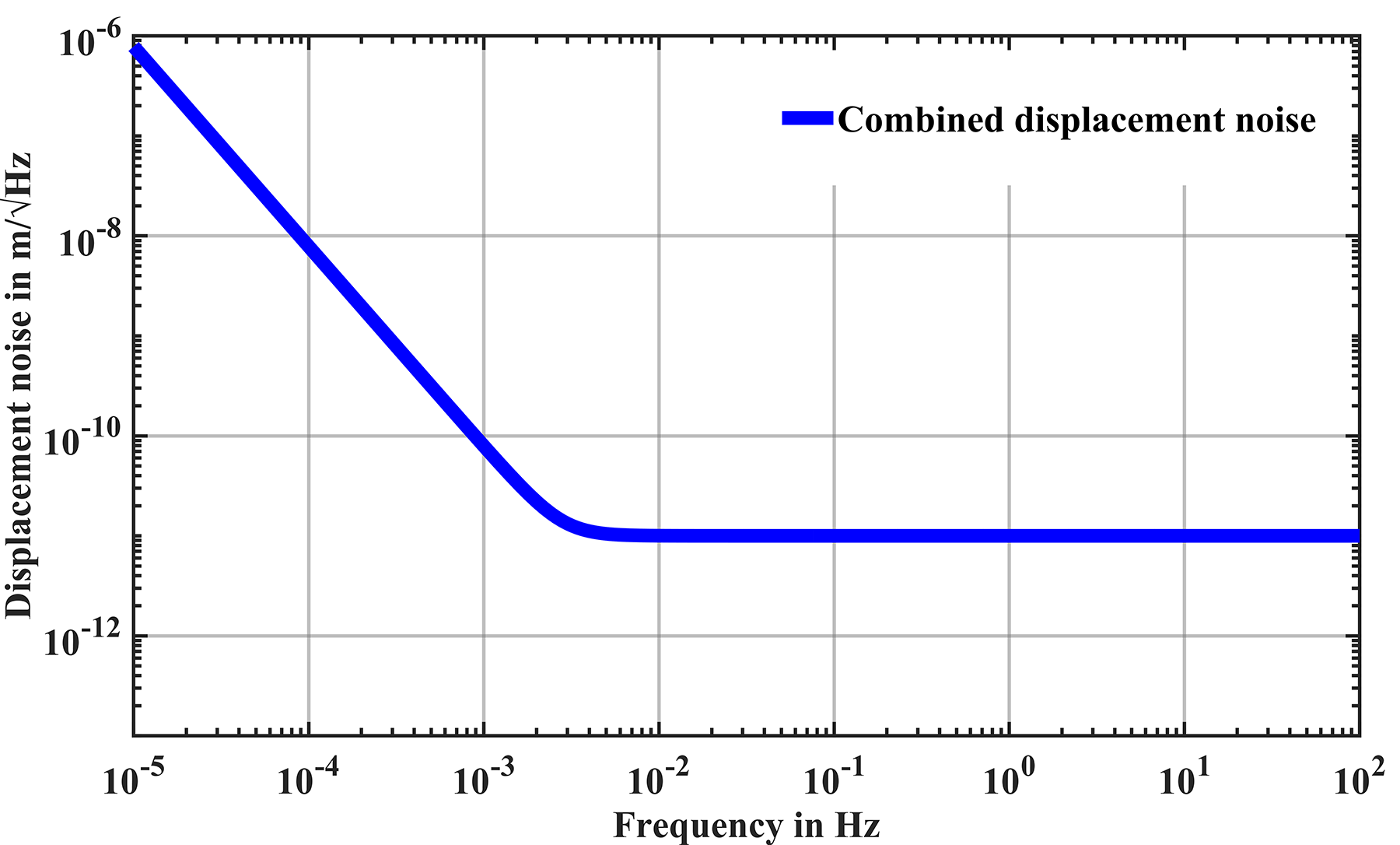}
		\caption{PAA noise model.} \label{noise_model}
	\end{figure}

	The distribution of noise in the time domain is obtained by backpropagating the noise expression in the frequency domain using the LTPDA toolbox in with the flowchart shown in Fig.~\ref{ltpda} \cite{ltpda}. Fig.~\ref{noise_model} is the PSD curve of noise. The total displacement noise PSD is a linear superposition of a number of individual noise PSDs.

	\clearpage
	
	\section{\label{sec8}Simulation results.}%1级标题
	
	\begin{figure}[hbt!]
		\centering
		\subfigure[PSD curves of the out-of-plane PAA between SC1 and SC2 before filtering and after filtering.]{
			\begin{minipage}{17cm} %[b]%{0.2\textwidth} 
				\includegraphics[width=0.8\textwidth,height=0.25\textwidth]{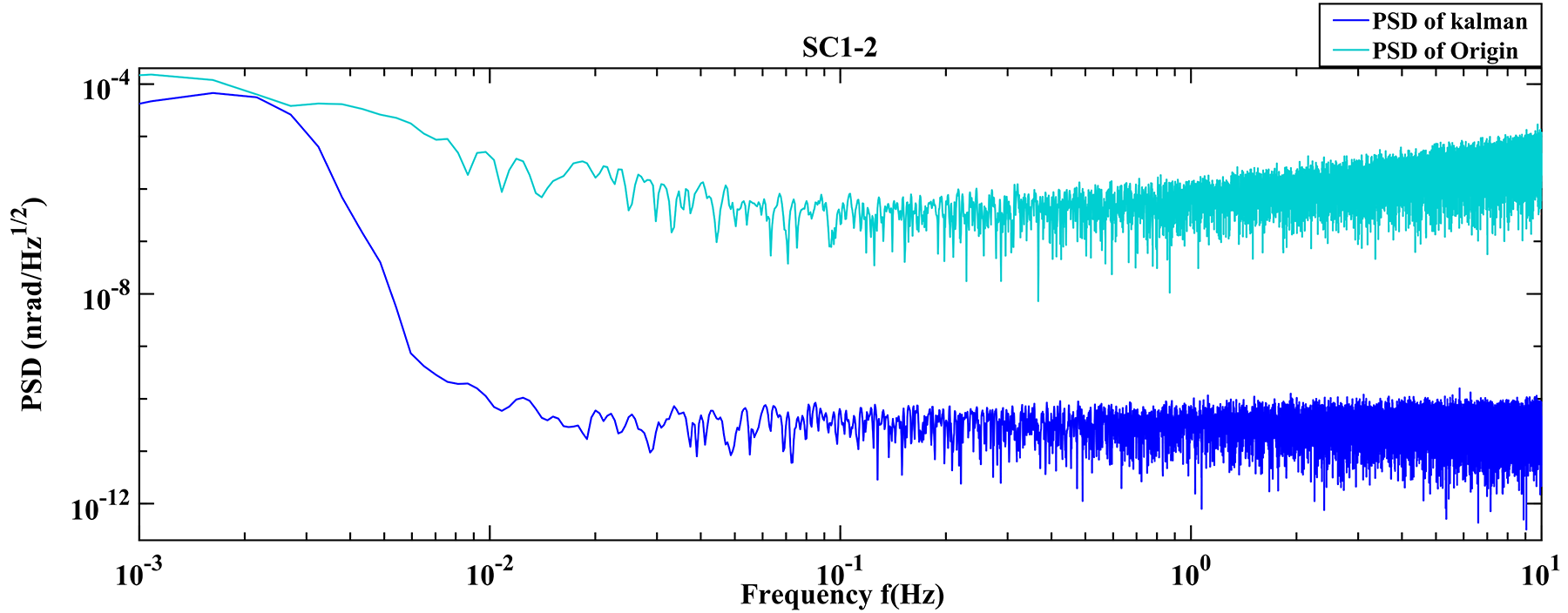} \\
				\label{PAAoutFrequency}
			\end{minipage}
		}
		\subfigure[PSD curves of the out-of-plane PAA prediction errors between SC1 and SC2 before filtering and after filtering.]{
			\begin{minipage}{17cm}
				\includegraphics[width=0.8\textwidth,height=0.25\textwidth]{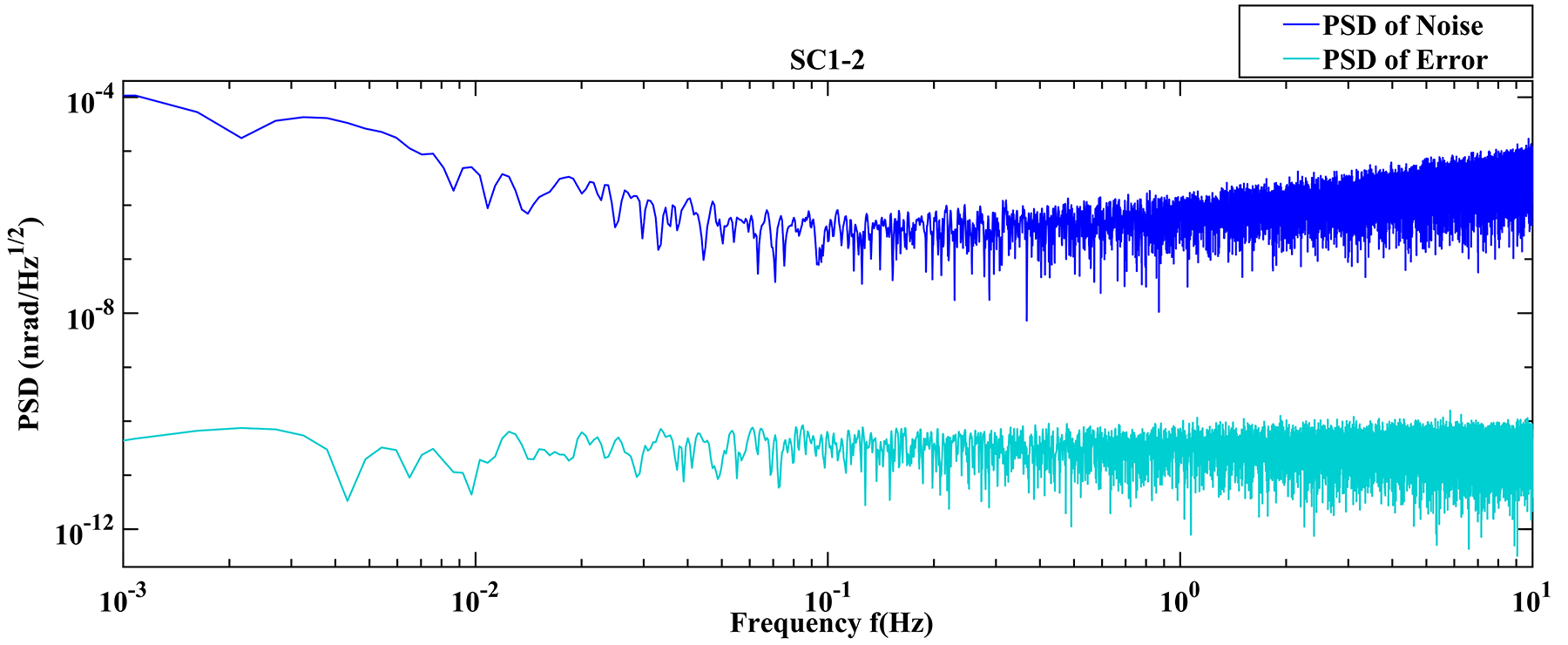} \\
				\label{PAAouterrorFrequency}	
			\end{minipage}
		}
		\caption{PSD curves of the out-of-plane PAA between SC1 and SC2. } 
	\end{figure}
	
	% fig12那一段
	Fig.~\ref{PAAoutFrequency} is the PSD curves of the out-of-plane PAA between SC1 and SC2 before filtering and after filtering. Fig.~\ref{PAAouterrorFrequency} is the PSD curves of the out-of-plane PAA errors between SC1 and SC2 before filtering and after filtering. Both figures show that the AEKF effectively filters high-frequency noise signals and low-frequency noise between 1 mHz and 10 Hz. The filtering effect of the AEKF in the high-frequency band is better than that at low frequency. In the band from 1 mHz to 0.01 Hz, the PSD of the out-of-plane PAA before filtering SC1 to SC2 is about $1.51\times 10^{-4}\; {\hbox{rad}}/\sqrt{\hbox{\hbox{Hz}}}$, and it is reduced to about $4.69\times 10^{-5}\; {\hbox{rad}}/\sqrt{\hbox{Hz}}$ after filtering. In the band from 1 mHz to 0.01 Hz, the PDS of the out-of-plane PAA before filtering is about $3.76\times 10^{-6}\; {\hbox{rad}}/\sqrt{\hbox{Hz}}$, and it is reduced to about $5.88\times 10^{-11}\;\hbox{rad}/\sqrt{\hbox{Hz}}$ after filtering. In the frequency window 1 Hz-10 Hz,  The AEKF's noise PSD rejection ratio is about -3.9 dB near 1mHz, -48 dB near 0.01Hz, and close to -39 dB at 0.01Hz-10Hz.
	
	\begin{figure}[hbt!]
		\centering
		\subfigure[PSD curves of the in-plane PAA between SC1 and SC2 before filtering and after filtering.]{
			\begin{minipage}{17cm} %[b]%{0.2\textwidth} 
				\includegraphics[width=0.8\textwidth,height=0.25\textwidth]{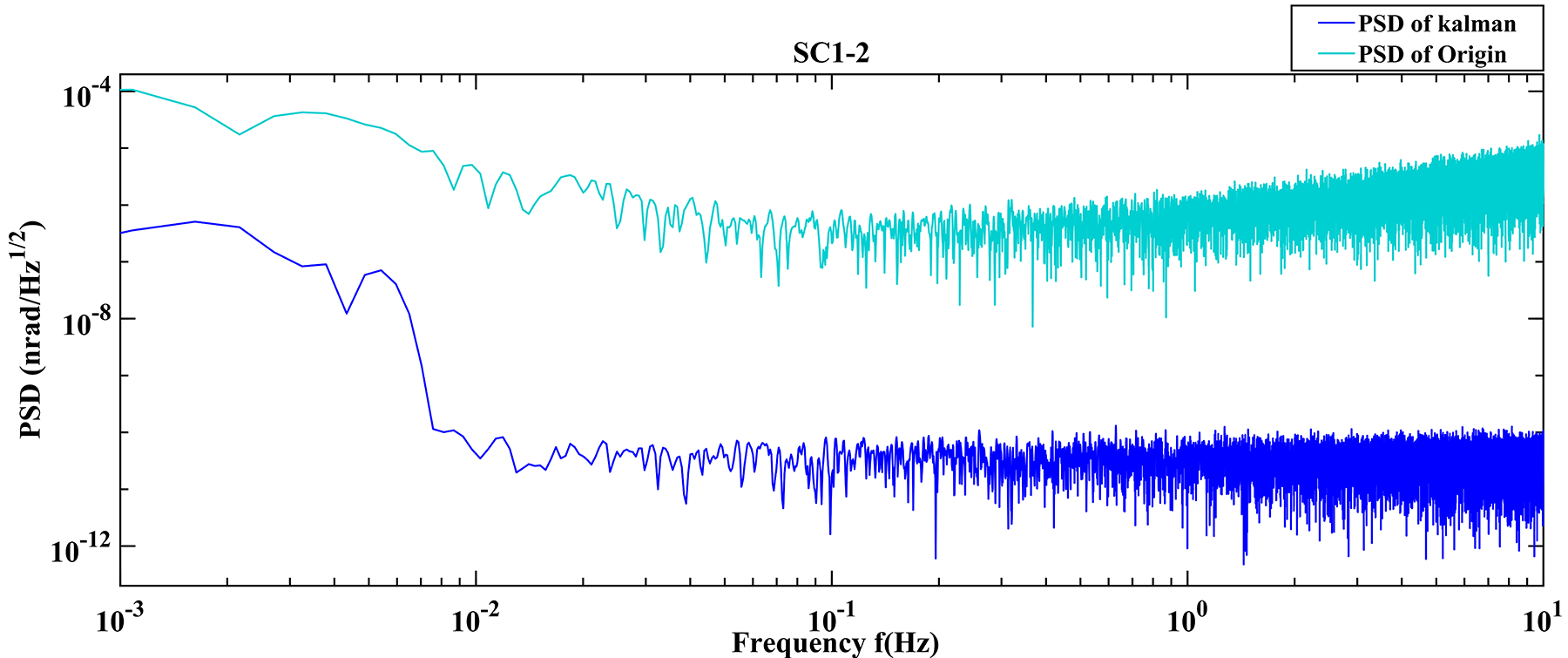} \\
				\label{PAAinFrequency}
			\end{minipage}
		}
		\subfigure[PSD curves of the in-plane PAA errors between SC1 and SC2 before filtering and after filtering.]{
			\begin{minipage}{17cm}
				\includegraphics[width=0.8\textwidth,height=0.25\textwidth]{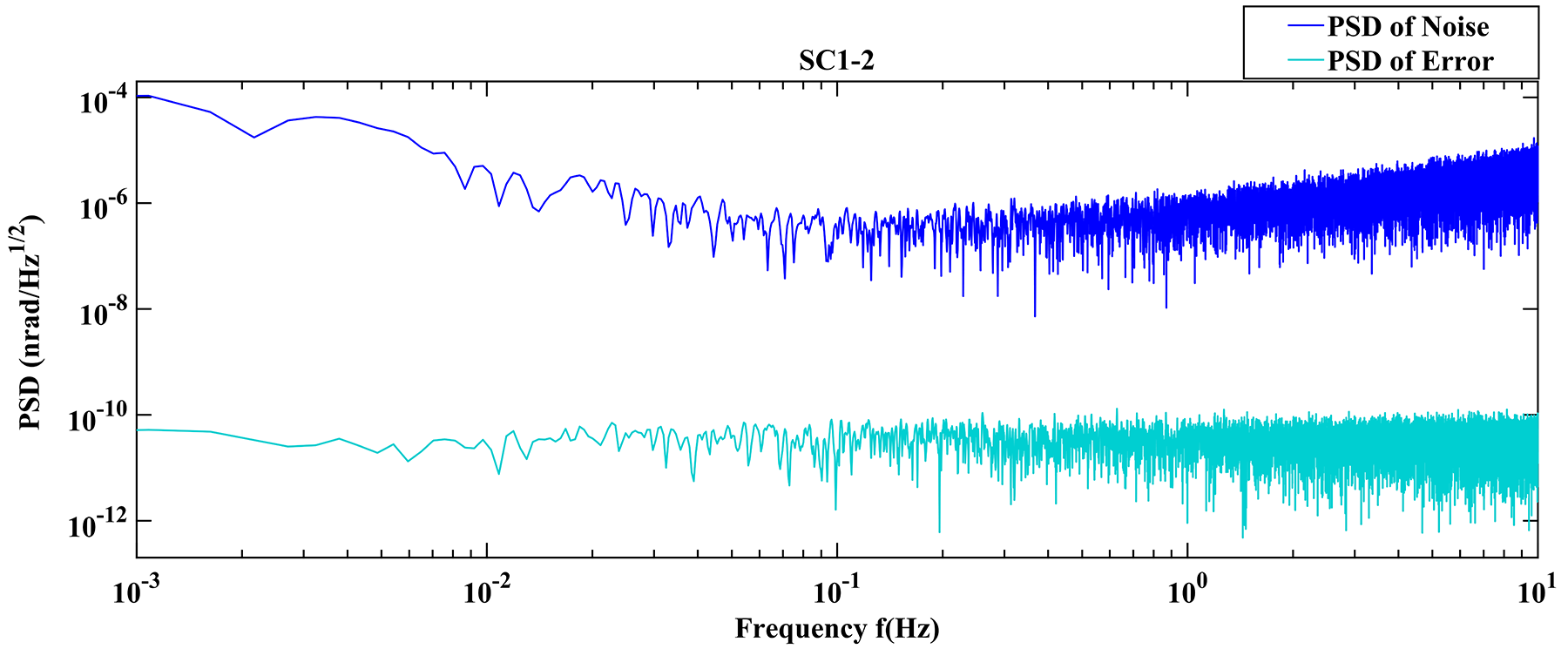} \\
				\label{PAAinerrorFrequency}	
			\end{minipage}
		}
		\caption{PSD curves of the in-plane PAA between SC1 and SC2 } 
	\end{figure}
	Fig.~\ref{PAAinFrequency} is the PSD curves of the in-plane PAA between SC1 and SC2 before filtering and after filtering. Fig.~\ref{PAAinerrorFrequency} is the PSD curves of the in-plane PAA errors between SC1 and SC2 before filtering and after filtering. Both Figures show that the AEKF effectively filters high-frequency noise signals and low-frequency noise between 1mHz and 10 Hz. The filtering effect of the AKEF in the high-frequency band is better than that at low frequency. In the band from 1mHz to 0.01Hz, the PSD of the in-plane PAA before filtering is about $1.06\times 10^{-4}\; \hbox{rad}/\sqrt{\hbox{Hz}}$, and it is reduced to about $3.57\times 10^{-7}\;\hbox{rad}/\sqrt{\hbox{\hbox{Hz}}}$ after filtering. In the band from 0.01 Hz to 10 Hz, the PSD of the in-plane PAA before filtering is about $3.76\times 10^{-6}\;\hbox{rad}/\sqrt{\hbox{Hz}}$, and it is reduced to about $3.48\times 10^{-11}\;\hbox{rad}/\sqrt{\hbox{Hz}}$ after filtering. The noise PSD rejection ratio of the AEKF is about -24.5dB near 1 mHz, -50.4dB near 0.01Hz, and close to -42dB at 0.01Hz-10Hz.
	
	\clearpage
	\begin{figure}[hbt!]
		\includegraphics[width=0.8\textwidth,height=0.25\textwidth]{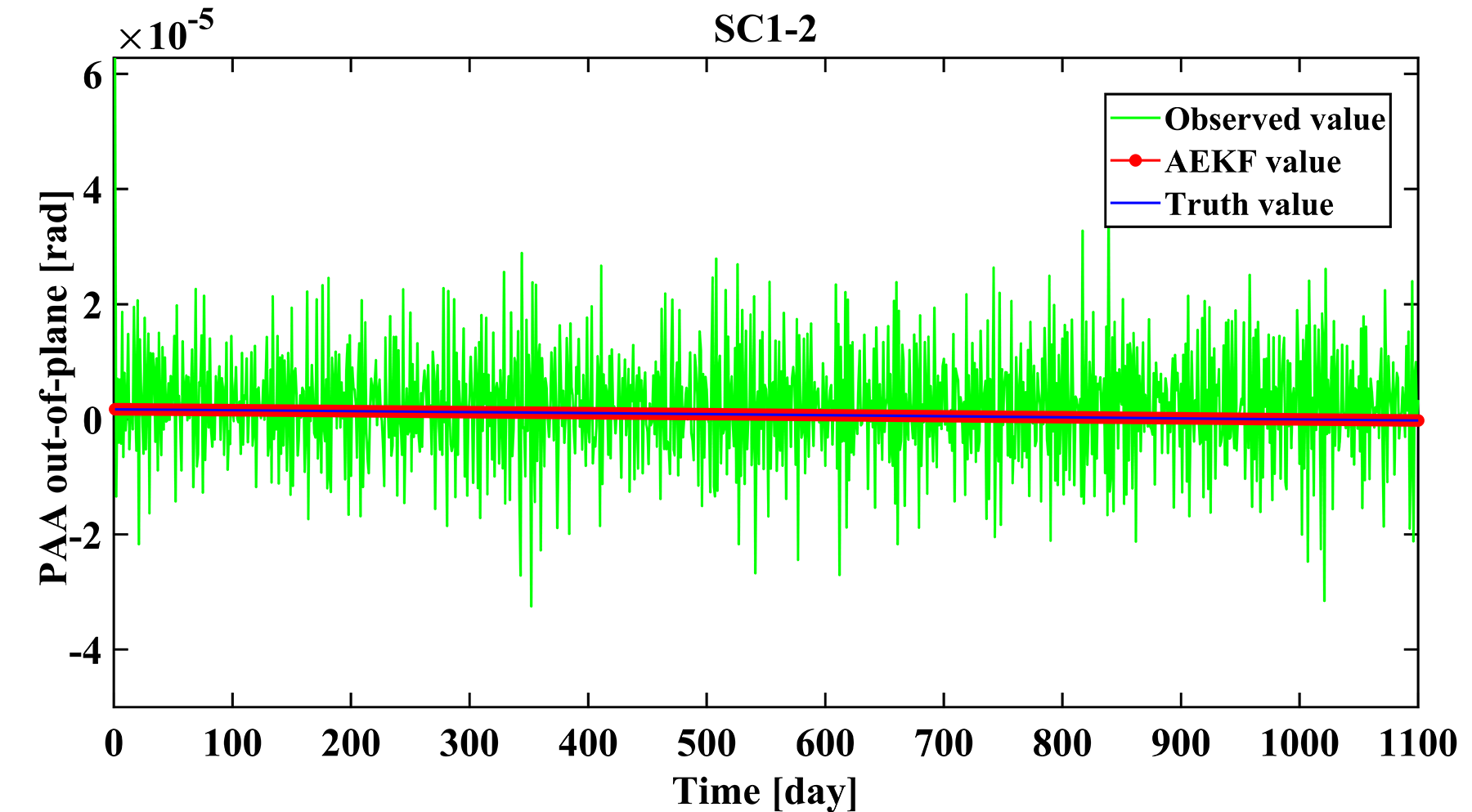}
		\caption{Out-of-Plane PAA between SC1 and SC2 before filtering and after filtering.} \label{PAAouttime}
	\end{figure}
	
	Fig.~\ref{PAAouttime} shows the time domain noise curve of the out-of-plane PAA between SC1 and SC2 before filtering and the curve after Kalman filtering. This result reflects the time domain performance of the AEKF. The results show that the AEKF is efficient in noise suppression.
	
	\begin{figure}[hbt!]
		\includegraphics[width=0.8\textwidth,height=0.25\textwidth]{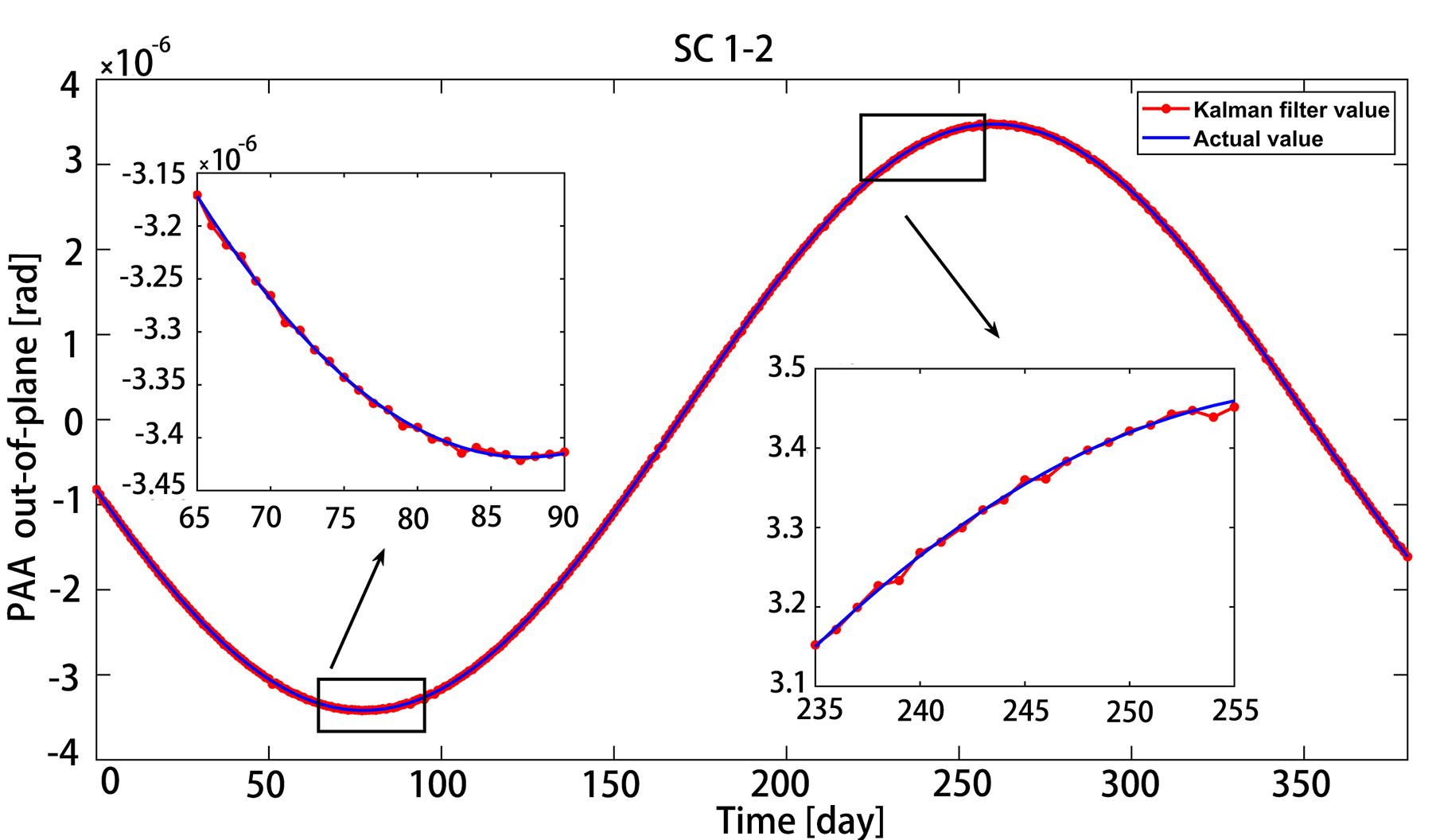}
		\caption{Out-of-plane PAA between SC1 and SC2 after filtering.} \label{PAAouttimeone}
	\end{figure}
	\begin{figure}[hbt!]
		\includegraphics[width=0.8\textwidth,height=0.25\textwidth]{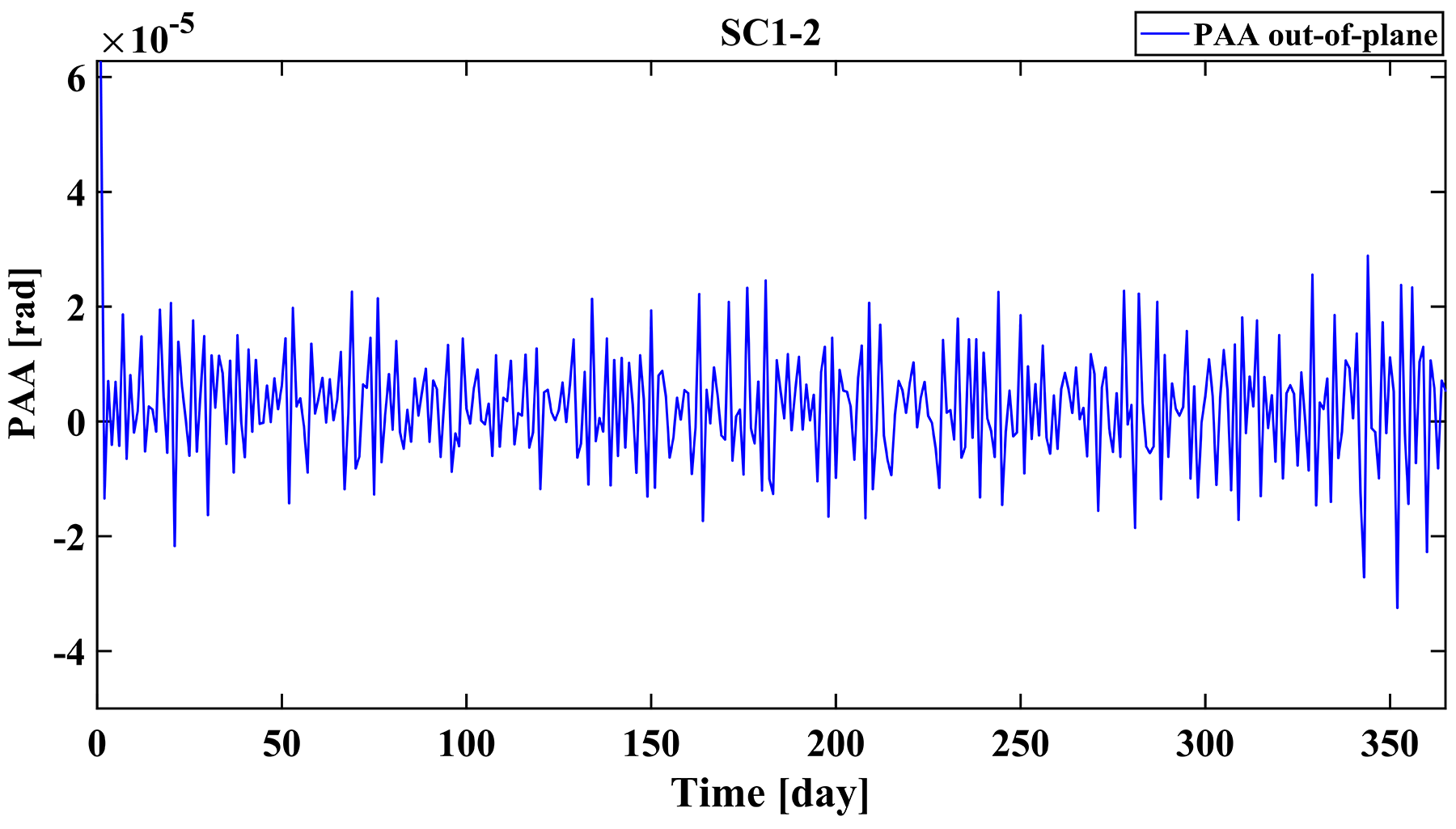}
		\caption{Prediction error of the out-of-plane PAA between SC1 and SC2.} \label{PAAouterrortime}
	\end{figure}
	%fig9
	Fig.~\ref{PAAouttimeone} displays the AEKF result and the real values of the out-of-plane PAA between SC1 and SC2 variation in one year. Fig.~\ref{PAAouterrortime} shows the prediction error of the out-of-plane PAA between SC1 and SC2 after Kalman filtering in the time domain. During 100-180 days and 300-360 days, their relative motion is uniform, and the Kalman filtered out-of-plane PAA closely matches the actual situation. At this time, the noise before filtering is about 63 $\mu \hbox{rad}$, and after Kalman filtering, it reaches about 0.84 $\hbox{nrad}$. However, within 50-100 days and 180-300 days, there is a significant variation in the out-of-plane PAA. Consequently, the filter error during these periods is relatively large, about 0.95 $\hbox{nrad}$. This is mainly due to the large variation of relative velocity motion between two SCs.
	
	\begin{figure}[hbt!]
		\includegraphics[width=0.8\textwidth,height=0.35\textwidth]{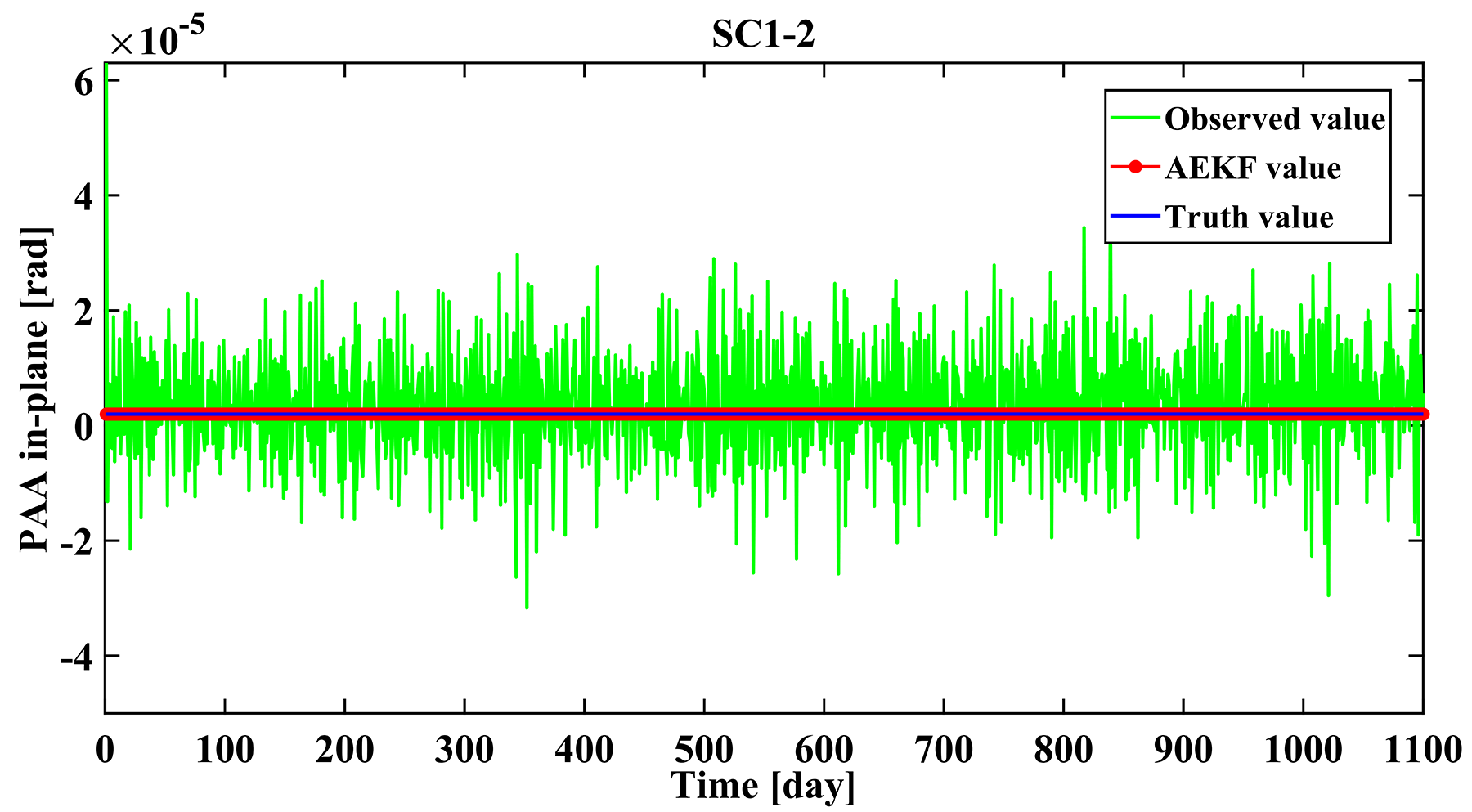}
		\caption{ In-plane PAA between SC1 and SC2 before filtering and after filtering .} \label{PAAintime}
	\end{figure}
	%fig10
	Fig.~\ref{PAAintime} displays the time domain noise curve of the in-plane PAA between SC1 and SC2 before filtering an after filtering. The results in the time domain reflect the noise reduction effect of the in-plane PAA under the AEKF. Compared with the out-of-plane PAA, the variation range of the in-plane PAA is reduced and the variation under the action of noise is evident. The in-plane PAA noise in the time domain shows a decreasing trend in the range of 50-100 days and an increasing trend in the range of 100-200 days. 
	\clearpage
	\begin{figure}[hbt!]
		\includegraphics[width=0.8\textwidth,height=0.3\textwidth]{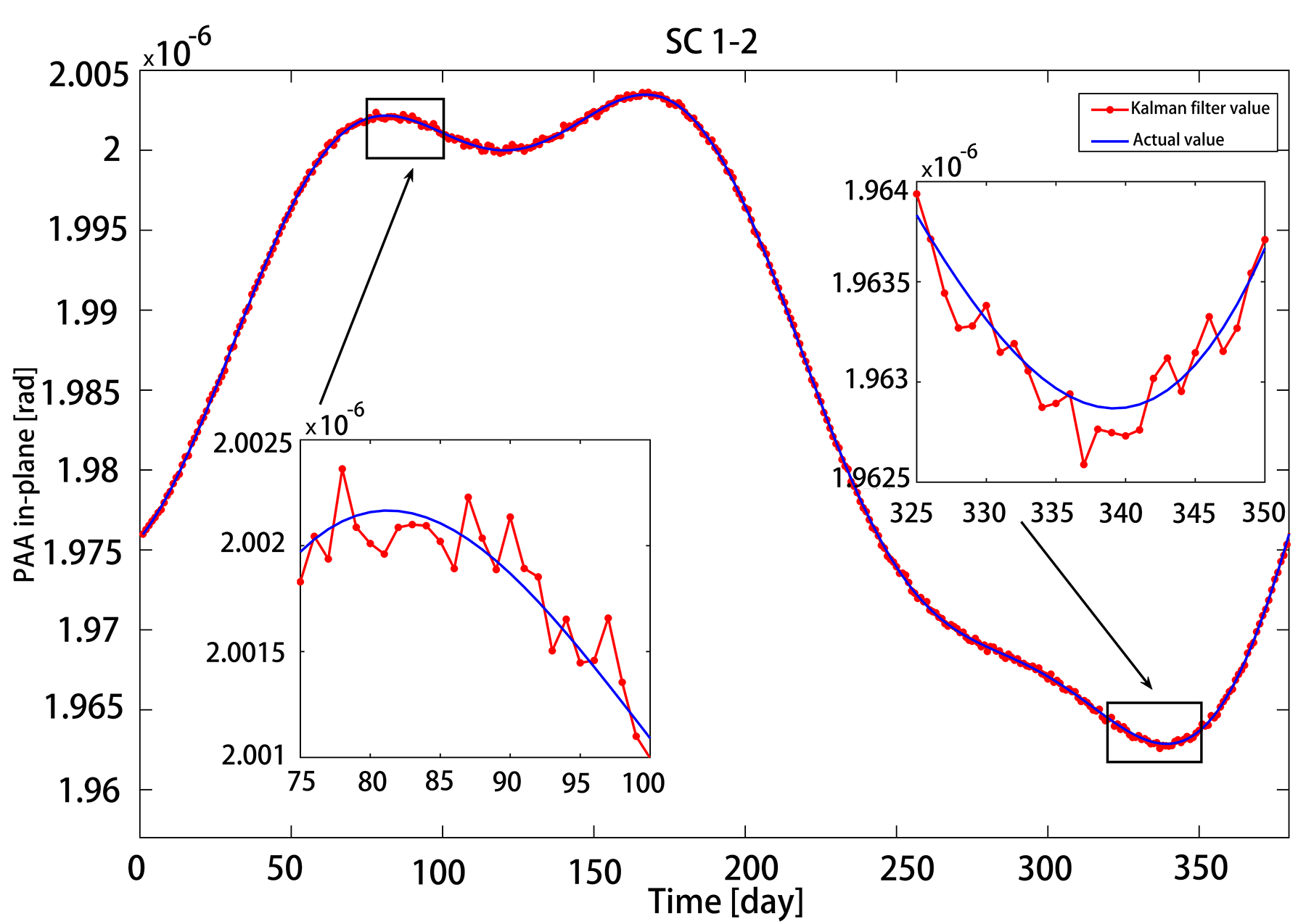}
		\caption{In-plane PAA between SC1 and SC2 after filtering.} \label{PAAintimeone}
	\end{figure}
	\begin{figure}[hbt!]
		\includegraphics[width=0.8\textwidth,height=0.30\textwidth]{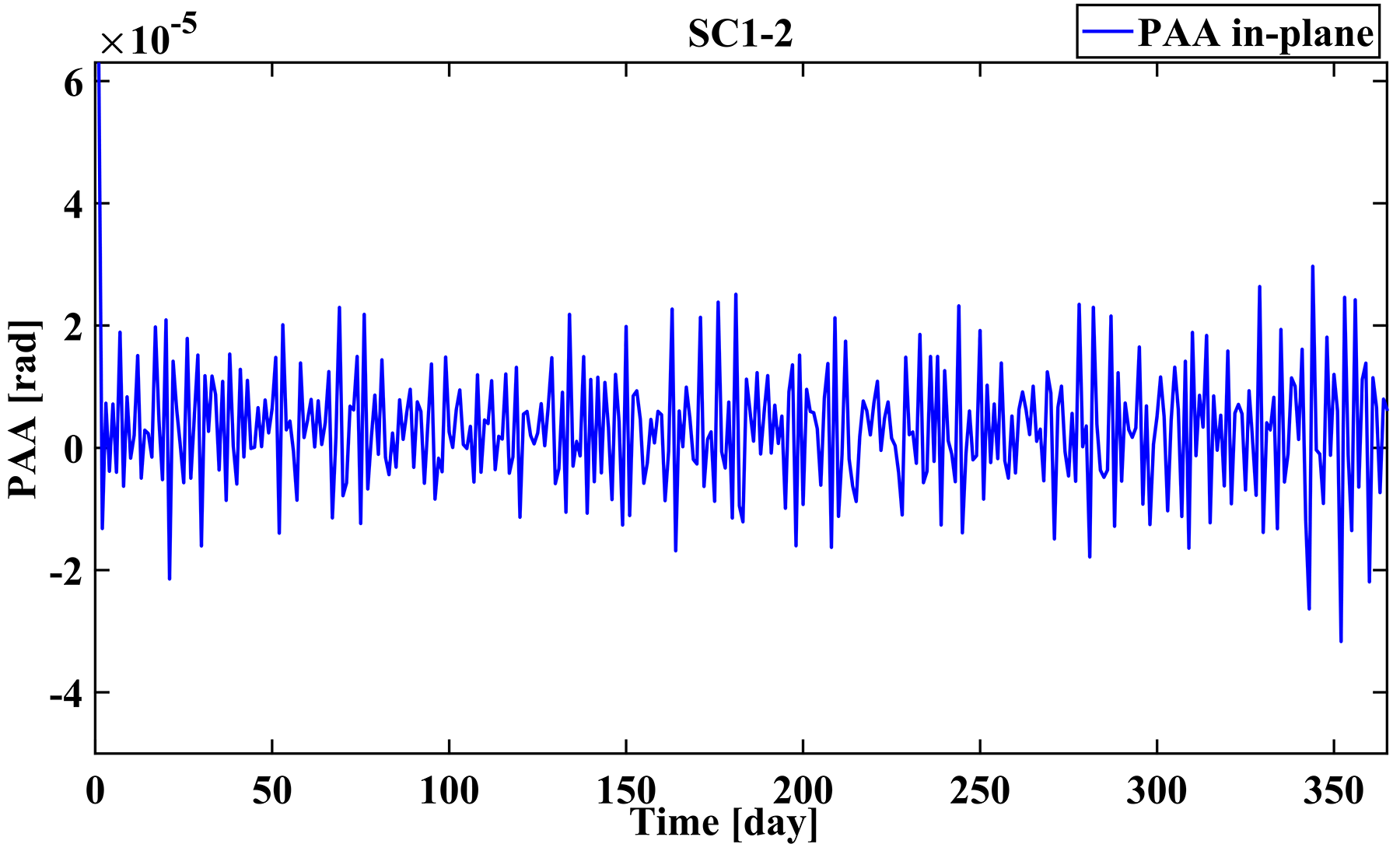}
		\caption{Prediction error of the in-plane PAA between SC1 and SC2.} \label{PAAinerrortime}
	\end{figure}
	% fig11那一段
	Fig.~\ref{PAAintimeone} shows the AEKF result and the real values of the in-plane PAA between SC1 and SC2 variation in one year. Fig.~\ref{PAAinerrortime} shows the prediction error of the in-plane PAA between SC1 and SC2 after Kalman filtering in the time domain. Compared with the out-of-plane PAA, the error variation of the in-plane PAA is more evident after Kalman filtering. Comparing the error before and after filtering,
	it may be seen that the AEKF algorithm can effectively suppress the in-plane PAA noise. The noise before filtering is about 63 $\mu \hbox{rad}$, while the pointing error after AEKF filtering is less than 1 $\hbox{nrad}$.
	\clearpage
	\subsection{Summary of the main results.}
	
	\begin{table}[htbp]
		\caption{\label{table1} Dynamic range of the PAA.}
		\centering
		\begin{ruledtabular}
			\begin{tabular}{ccccccc}
				Parameters & SC12 & SC13 & SC21 & SC23 & SC31 & SC32 \\ [0.5ex] 
				\hline
				Observed PAA out/(rad)& 6.987E-5  &	6.883E-5  &	6.916E-5	& 6.884E-5	& 6.812E-5	& 6.912E-5 \\
				Observed PAA in/(rad)& 6.584E-5 &	6.413E-5 &	6.622E-5 &	6.418E-5 &	6.602E-5 &	6.512E-5 \\	
				AEKF PAA out/(rad)& 6.899E-6 &	6.905E-6 &	6.899E-6 &	6.893E-6	& 6.905E-6 &	6.892E-6\\
				AEKF PAA in/(rad)& 3.743E-8	& 3.967E-8 &	3.750E-8 &	4.112E-8 &	4.65E-8 &	4.531E-8\\	
				Ideal PAA out/(rad)& 6.899E-6 &	6.905E-6 &	6.899E-6 &	6.892E-6 &	6.905E-6 &	6.892E-6\\
				Ideal PAA in/(rad)& 3.708E-8 &	6.633E-8 &	3.708E-8 &	4.483E-8 &	6.633E-8 &	4.483E-8\\	
			\end{tabular}
		\end{ruledtabular}
	\end{table}
	
	We will give a recap of the main results at the end of this section.  The dynamic range of PAA  is dependent on error in the precision orbit determination of SCs. When the position determination error of the SC is 20km and the velocity error is 2cm/s, the dynamic range of the out-of-plane PAA and the in-plane PAA is about 63 $\mu \hbox{rad}$.   With AEKF inserted into the control loop, the dynamic range of the out-of-plane PAA is approximately reduced to 6.9 $\mu \hbox{rad}$, and the dynamic range of the in-plane PAA is approximately reduced to 40 $\hbox{nrad}$.  The result shows that the dynamic range of the PAA can be effectively reduced by AEKF even in the presence of error in precision orbit determination.   
	
	Based on the analysis of the filtering results in the frequency domain, the error of the in-plane PAA and the out-of-plane PAA decreases significantly at high-frequency ends.  In the band from 0.01 Hz to 10 Hz, the filtering error signal is less than 1  ${\hbox{nrad}}/\sqrt{\hbox{Hz}}$. The filtering effect of AEKF at low frequency is not as good as that at high frequency.

	\begin{table}[htbp]
		\caption{\label{table2}Prediction error of PAA in-plane and out-of-plane.}
		\centering
		\begin{ruledtabular}
			\begin{tabular}{ccccccc}
				Parameters & SC12 & SC13 & SC21 & SC23 & SC31 & SC32 \\ [0.5ex] 
				\hline
				PAA-OUT/(nrad)& 0.957 & 0.857 & 0.940 &0.941\
				& 0.882 & 0.852  \\
				PAA-IN(nrad)& 0.846 & 0.833 & 0.845 &0.920\
				& 0.972 & 0.814  \\	
			\end{tabular}
		\end{ruledtabular}
	\end{table}
	In our simulation, orbital dynamics and the update of the SC's state of motion are input into the iterations of the AEKF.
	Table \ref{table2} summarizes the results of the maximum prediction errors before and after filtering the in-plane PAA and the out-of-plane PAA between the two SCs. The results in Table \ref{table1} show that the AEKF can effectively suppress the pointing noise of the PAA.

	\begin{table}[htbp]
		\caption{\label{table3}Fitting result.}
		\centering
		\begin{ruledtabular}
			\begin{tabular}{ccccccc}
				Parameters & SC12 & SC13 & SC21 & SC23 & SC31 & SC32 \\ [0.5ex] 
				\hline
				SSE& 4.0985E-29 & 7.06114E-29 & 1.5231E-28 &7.4431E-29\
				& 5.5912E-29 & 8.10921E-30  \\
				RMSE& 1.7019E-16 & 2.23388E-16 & 3.28085E-16 &2.2935E-16\
				& 1.98781E-16 & 7.57027E-17  \\
				R-square& 1 & 1 & 1 &1\
				& 1 & 1  \\
				adjusted R-square& 1 & 1 & 1 &1\
				& 1 & 1  \\
			\end{tabular}
		\end{ruledtabular}
	\end{table}
	
	In our simulation, we conduct a fitting precision analysis on the discrete PAA values after filtering through the AEKF. The fitting accuracy can be evaluated using SSE, R-square, adjusted R-square, and Root Mean Squared Error (RMSE). The fitting results are shown in Table \ref{table3}. Both R-square and adjusted R-square are close to 1, SSE and RMSE are close to 0, indicating a good fit and that the fitting accuracy meets our requirements.

	\section{\label{sec9}Concluding remarks}
	
	In the present work, by exploiting the simple orbital structure of SC motion in the triangular constellation,  the feasibility of incorporating an AEKF in the PAAM is considered. The simulation results show that the dynamic range of the PAAM may be drastically reduced. This will cut down on the TTL coupling noise and the position noise budget allocated to the instrument. At the same time, the dynamic range of the fine pointing control is also reduced and leaves room for the improvement of accuracy for the fine pointing control. The position noise budget allocated to the pointing stability control may be reduced as a whole. 
	
	Though our calculations look complicated as we try to spell out the details of the derivation, the actual implementation of the Kalman filter is very simple, 
	From our flight experience with the adaptive AEKF for star sensors and, again, due to the simplicity of the orbital dynamics of the three SCs, we expect the  AEKF for PAAM will only consume a little computational resource on board. In our next step, we are setting up a desktop experiment to verify the AEKF constructed here in a hardware setting. Further, we are also considering the employment of the Kalman-filtered PAAM  to improve the capture time and efficiency in the ATP phase. We hope to report further progress in this direction soon.

	\begin{acknowledgments}
		This work is supported by the National Key R$\&$D Program of China (2022YFC2203701).
	\end{acknowledgments}

	\section{Appendices}
	
	In the main text, simulation results are presented only for a single arm out of the three arms in  the triangular constellation of SCs, as simulation results are almost identical for the the three arms. In this appendix, for completeness, simulation results for all three arms are presented. 

	\begin{figure}[hbt!]
		\includegraphics[width=0.8\textwidth,height=0.35\textwidth]{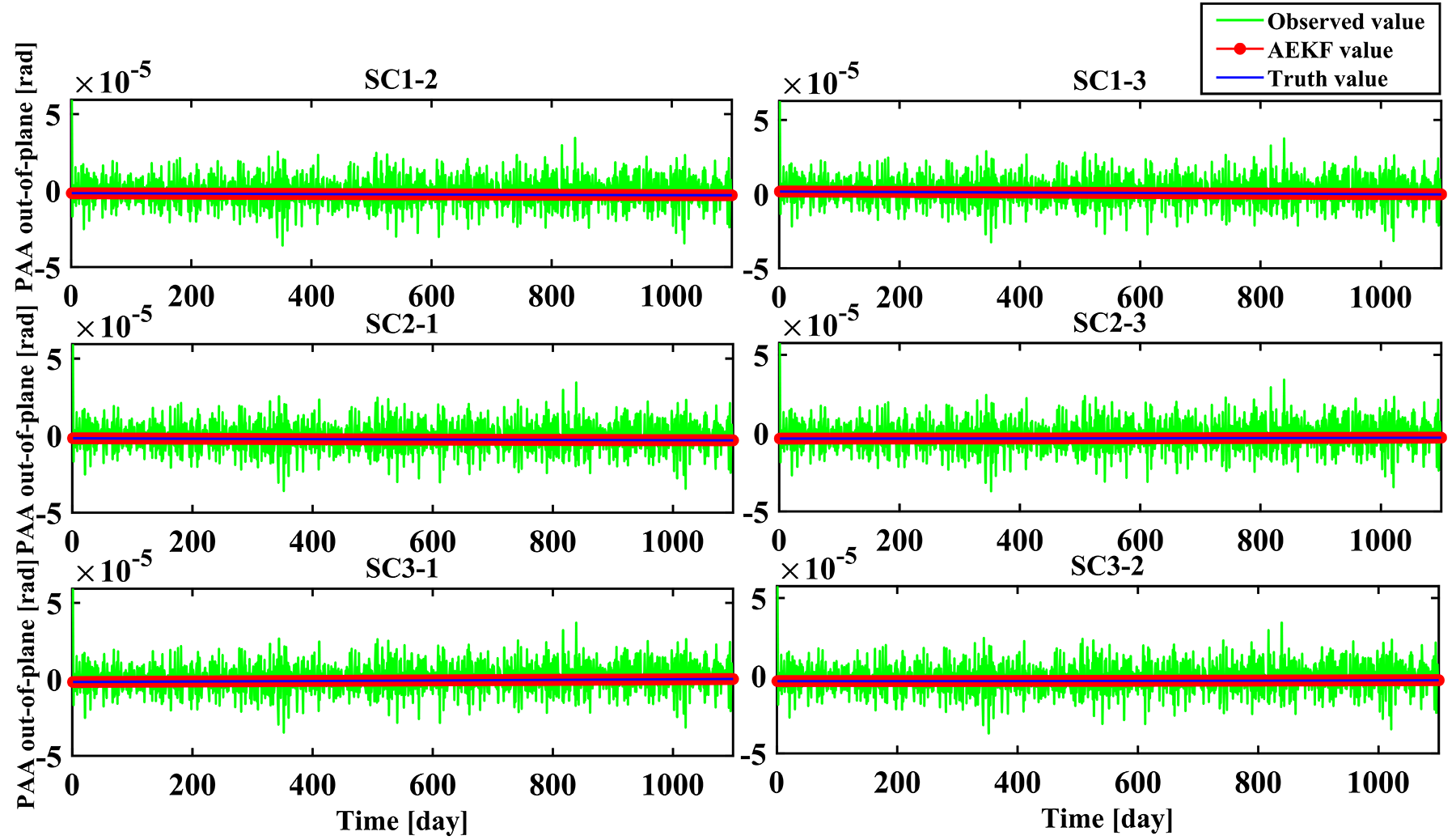}
		\caption{Out-of-plane PAA after filtering.} 
	\end{figure}
	
	\begin{figure}[hbt!]
		\includegraphics[width=0.8\textwidth,height=0.35\textwidth]{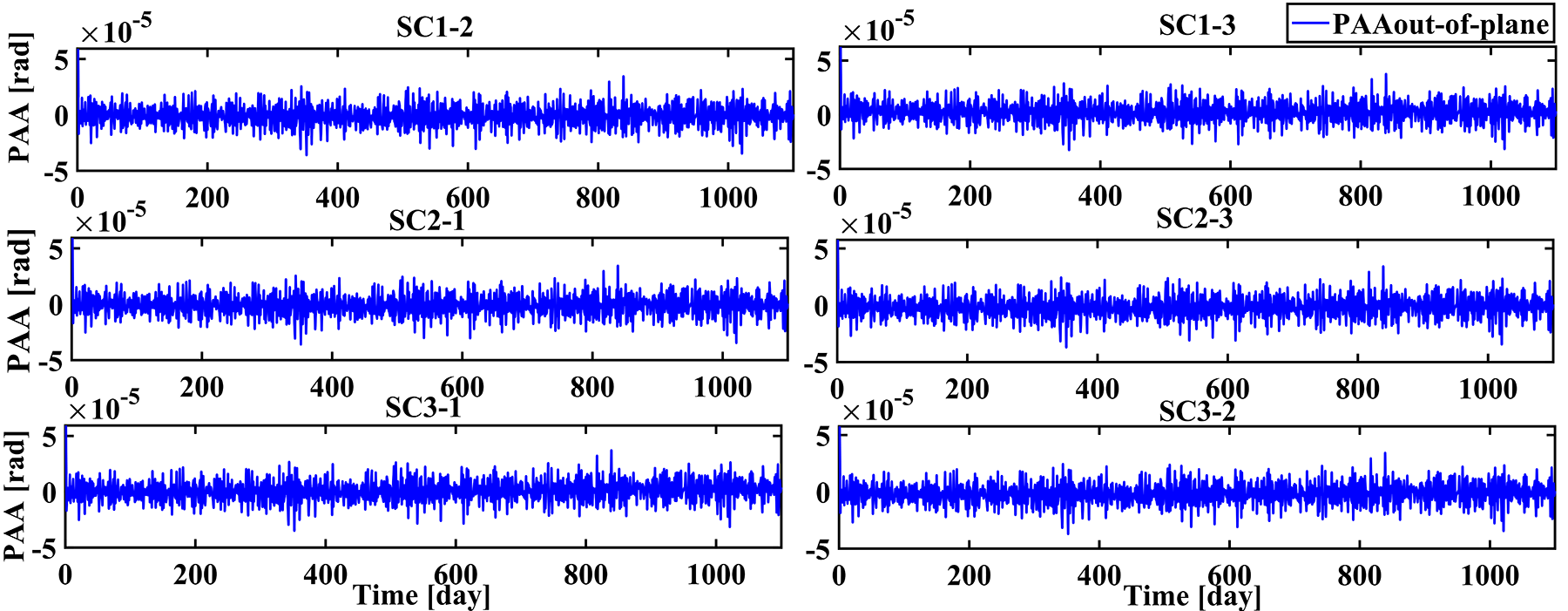}
		\caption{Prediction error of the out-of-plane PAA.} 
	\end{figure}
	
	\begin{figure}[hbt!]
		\includegraphics[width=0.8\textwidth,height=0.35\textwidth]{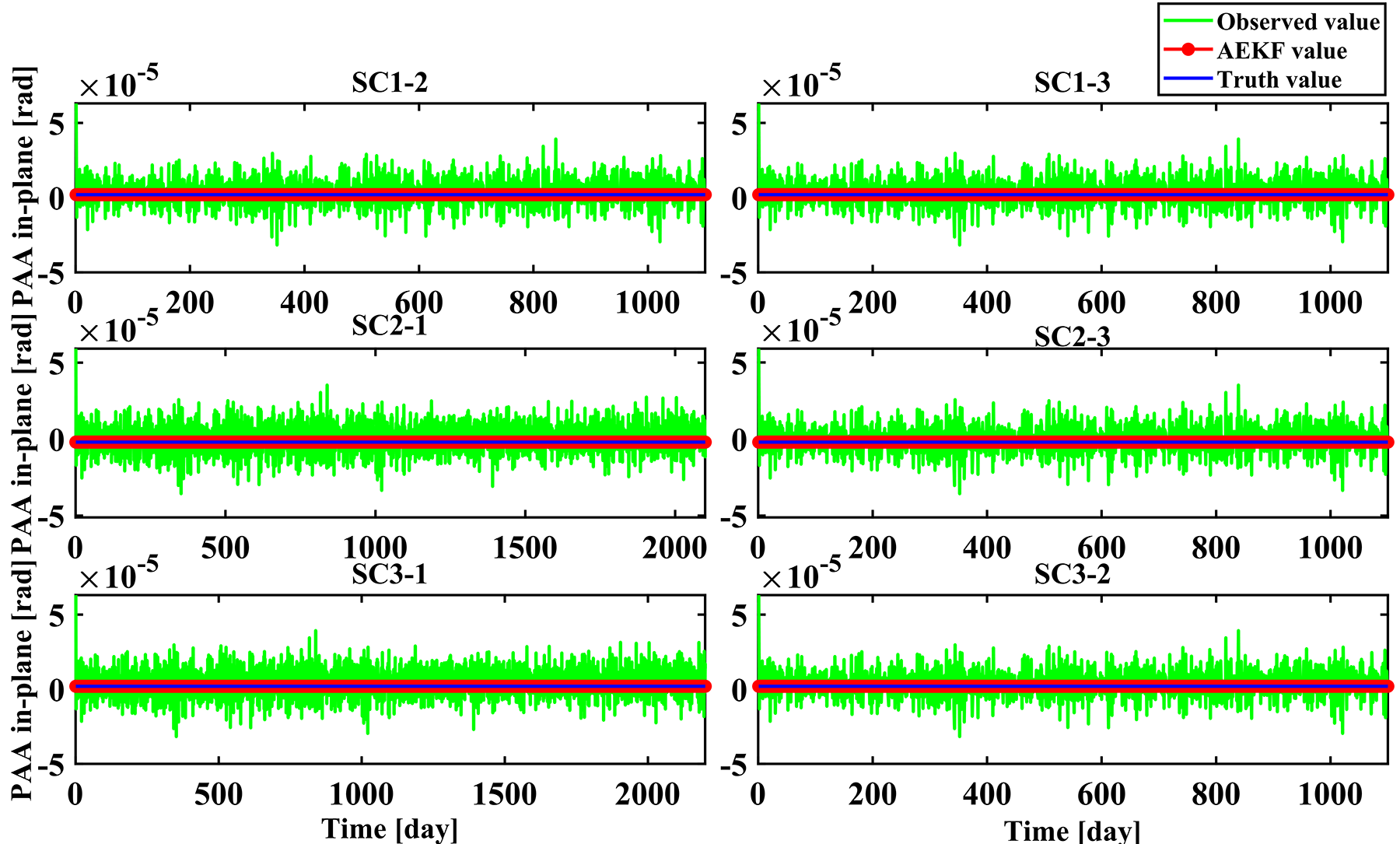}
		\caption{In-plane PAA after filtering.} 
	\end{figure}
	
	\begin{figure}[hbt!]
		\includegraphics[width=0.8\textwidth,height=0.35\textwidth]{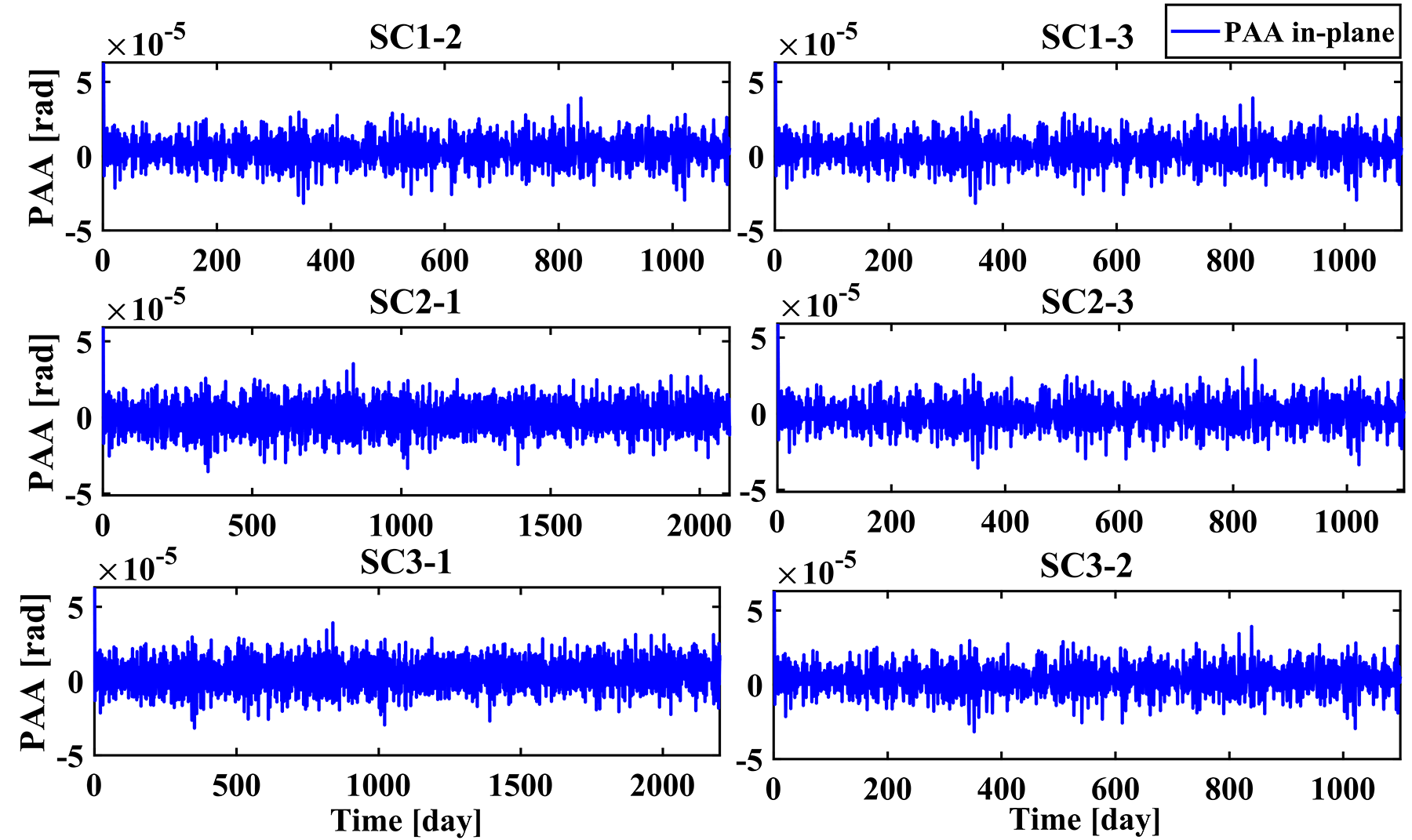}
		\caption{Prediction error of the in-plane PAA.} 
	\end{figure}
	
	\begin{figure}[hbt!]
		\includegraphics[width=0.8\textwidth,height=0.35\textwidth]{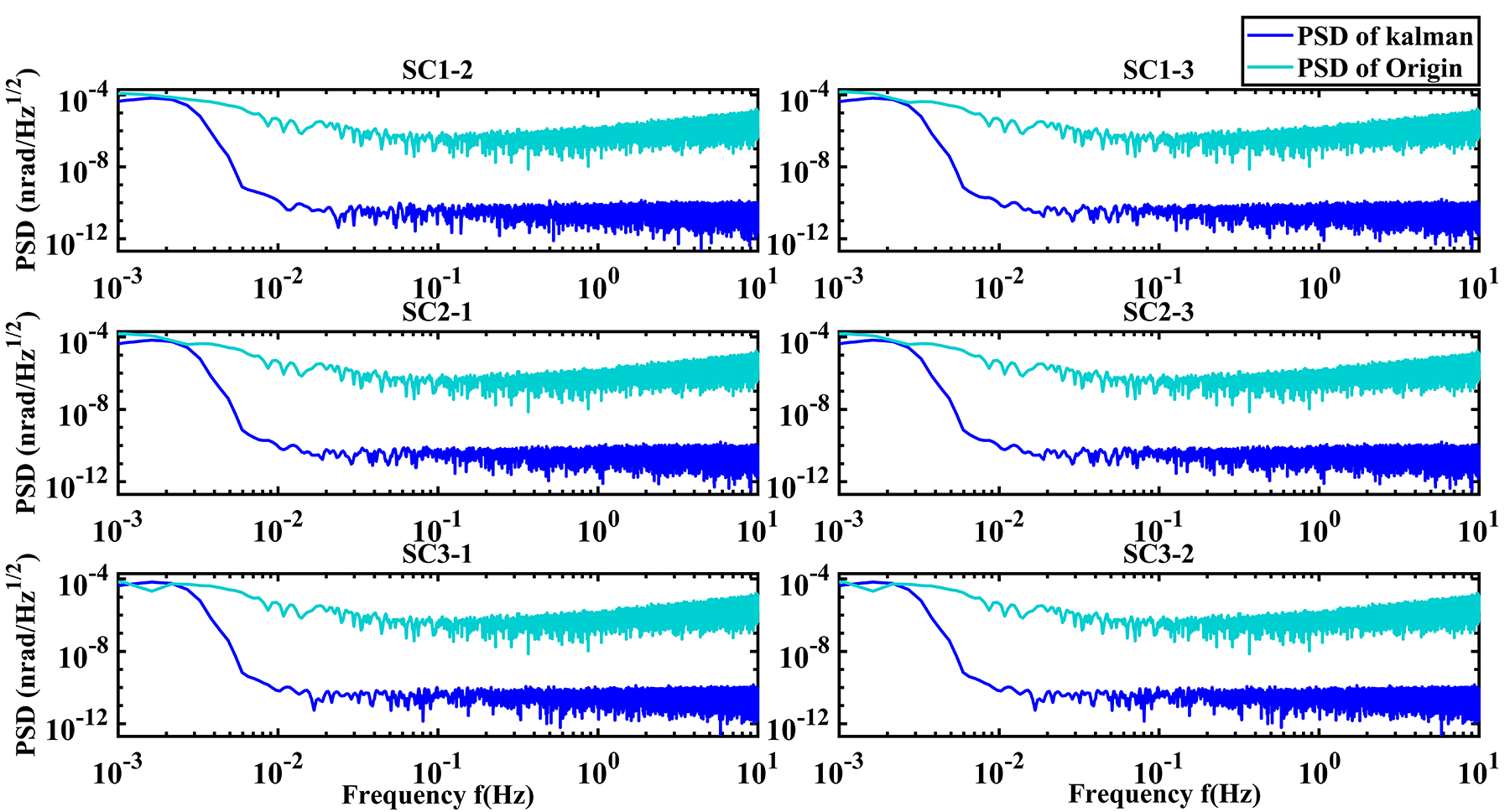}
		\caption{Out-of-plane PAA PSD curve before and after filtering.}
	\end{figure}
	
	\begin{figure}[hbt!]
		\includegraphics[width=0.8\textwidth,height=0.35\textwidth]{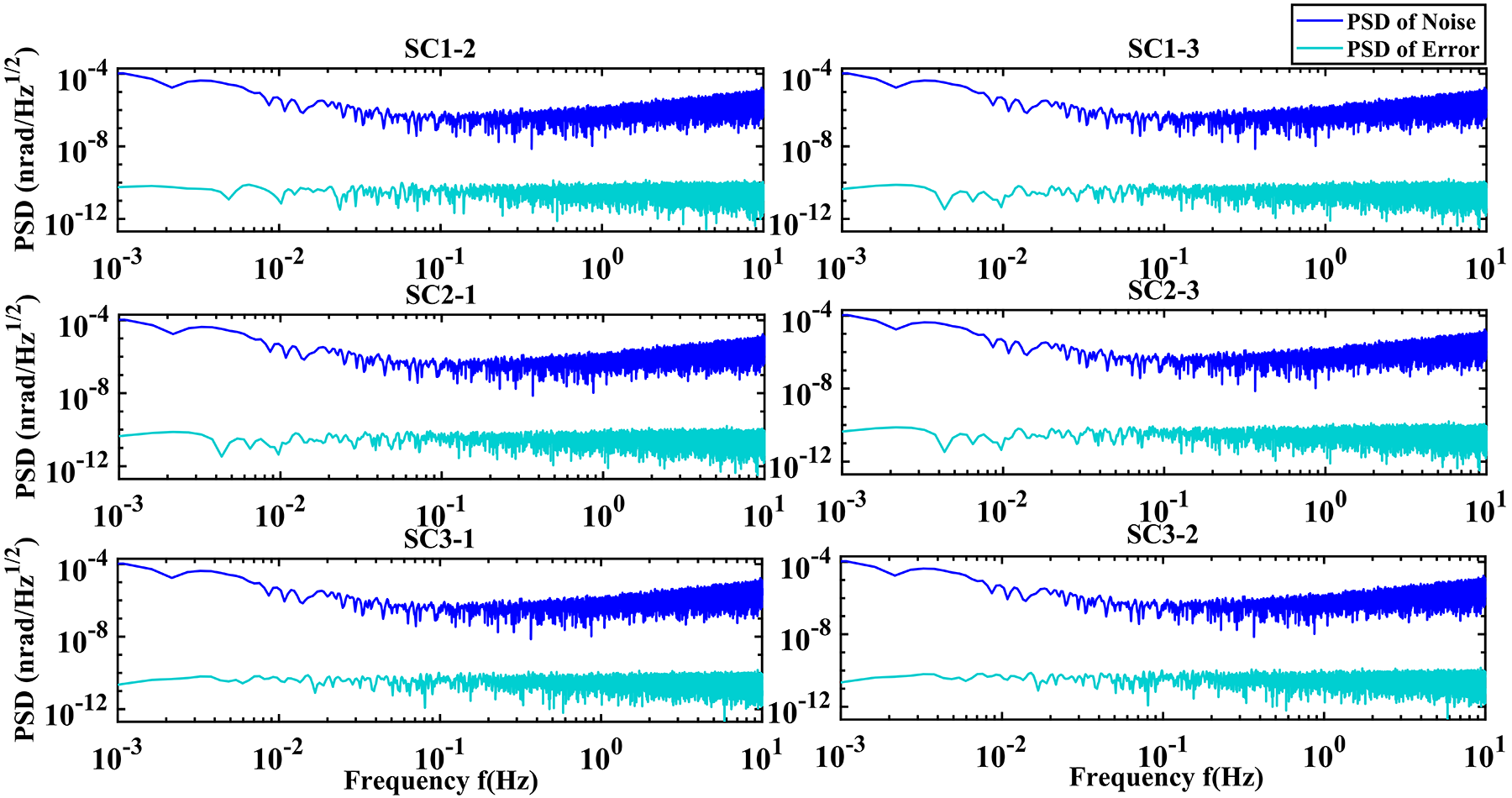}
		\caption{PSD curve of the out-of-plane PAA error before and after filtering.}
	\end{figure}
	
	\begin{figure}[hbt!]
		\includegraphics[width=0.8\textwidth,height=0.35\textwidth]{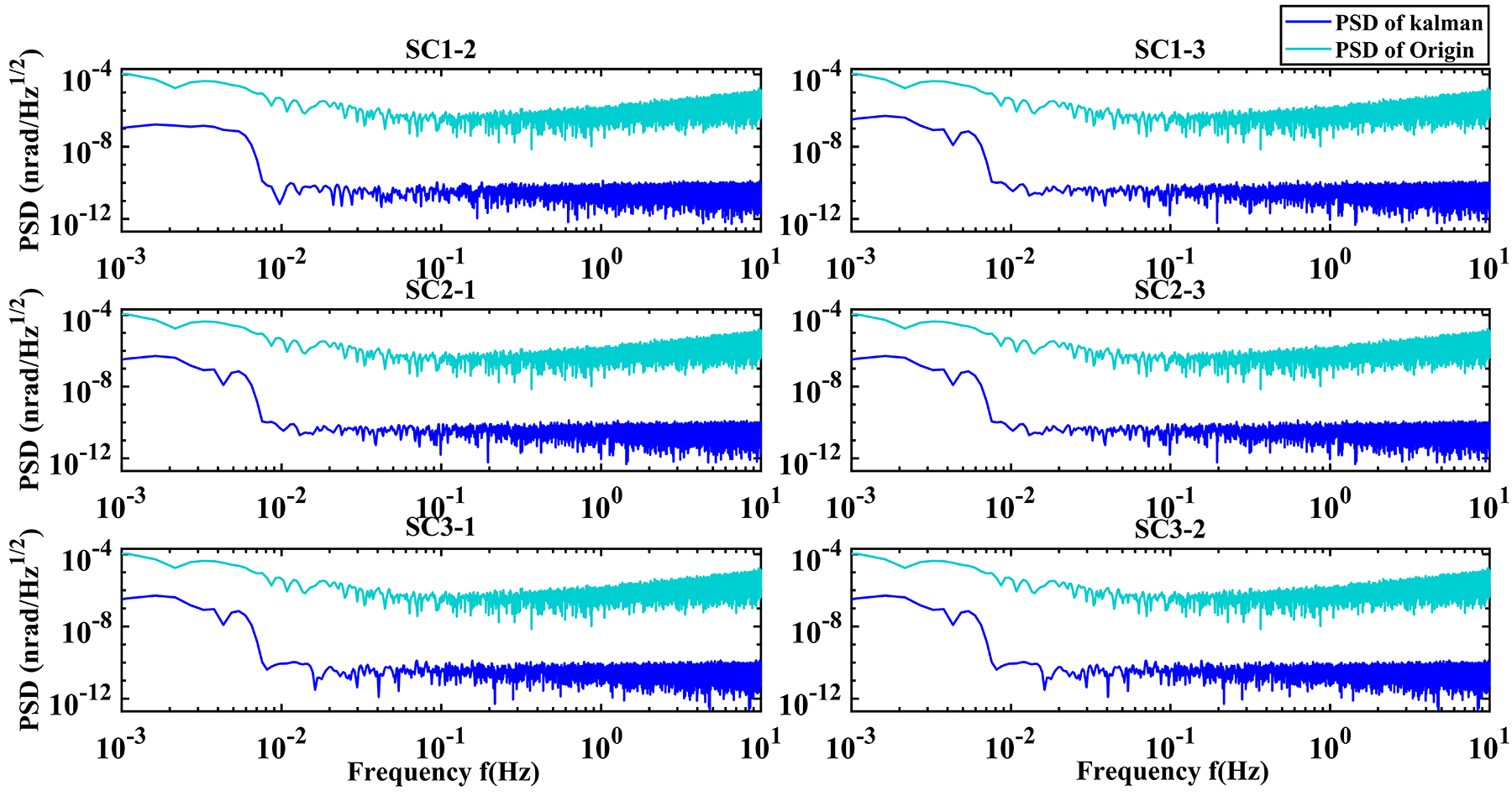}
		\caption{In-plane PAA PSD curve before and after filtering.}
	\end{figure}
	
	\begin{figure}[hbt!]
		\includegraphics[width=0.8\textwidth,height=0.35\textwidth]{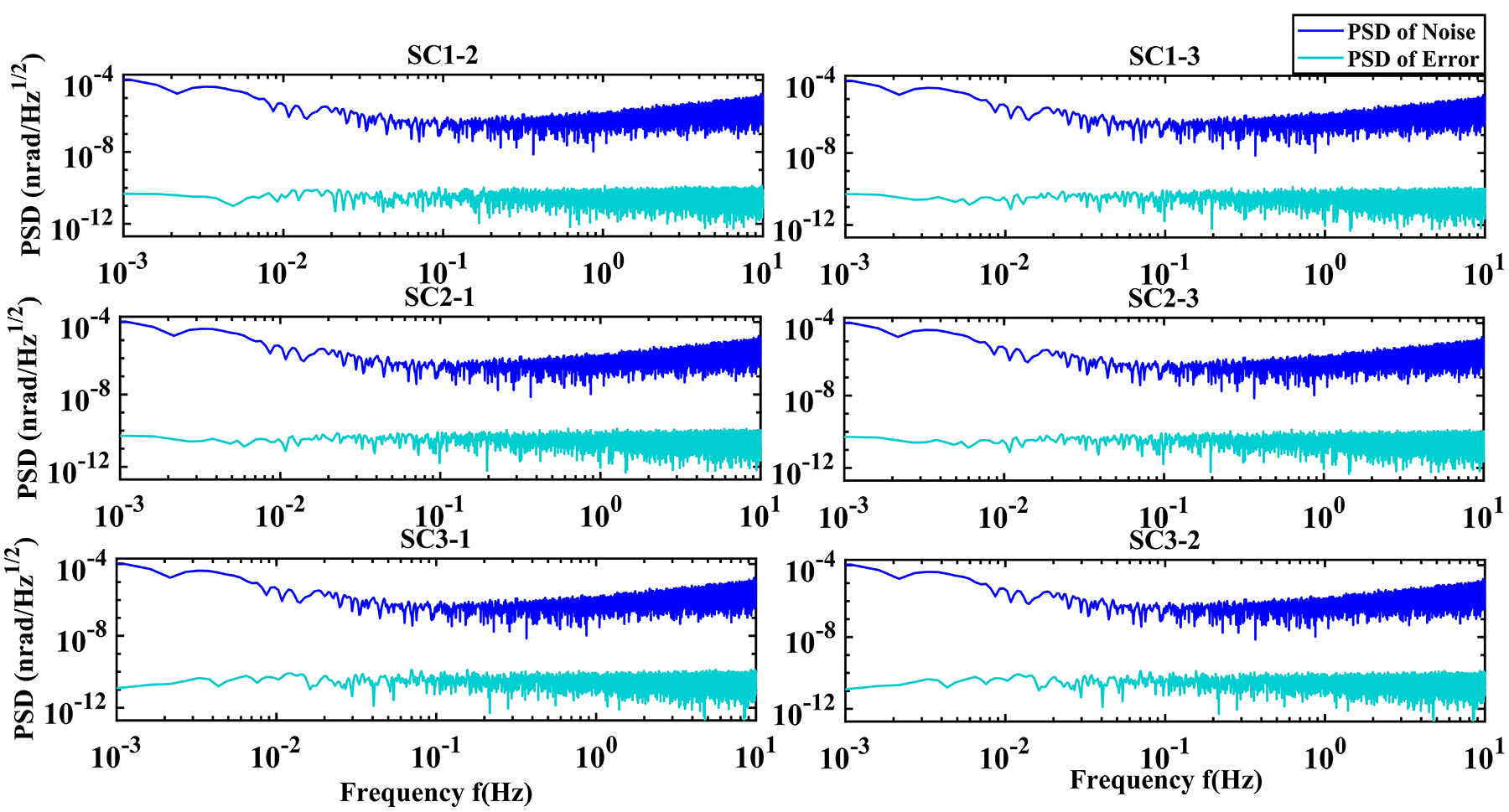}
		\caption{PSD curve of the in-plane PAA error before and after filtering.}
	\end{figure}

	\clearpage
	\newpage
	\vspace{155mm}
	
	\nocite{*}
	\bibliography{aps}% Produces the bibliography via BibTeX.
	
\end{document}